\documentclass[]{article}
\pdfoutput=1
\usepackage{amsmath,amssymb,amsfonts}
\usepackage{nicefrac}       
\usepackage[english]{babel}
\usepackage{microtype}
\usepackage{algorithmic}
\usepackage{graphicx}
\usepackage{textcomp}
\usepackage{booktabs}
\usepackage{color}
\usepackage{xcolor}
\usepackage{listings}
\usepackage[T1]{fontenc}
\usepackage{lmodern}
\usepackage{fancyvrb}
\usepackage[utf8]{inputenc}
\usepackage{mathtools}
\PassOptionsToPackage{hyphens}{url}\usepackage[hidelinks]{hyperref}
\usepackage{refcount}
\usepackage[autolanguage]{numprint}
\usepackage{subcaption}
\usepackage{rotating}
\usepackage{doi}
\usepackage{arxiv}

\colorlet{punct}{red!60!black}
\definecolor{background}{HTML}{EEEEEE}
\definecolor{delim}{RGB}{20,105,176}
\colorlet{numb}{magenta!60!black}

\VerbatimFootnotes

\def\uschema{{U-Schema}}

\def\urule#1{\noindent\textbf{#1}}

\lstdefinelanguage{cypher}{
    basicstyle=\scriptsize\ttfamily,
    numbers=none,
    numberstyle=\scriptsize,
    stepnumber=1,
    numbersep=8pt,
    showstringspaces=false,
    breaklines=true,
    frame=lines,
    backgroundcolor=\color{background},
    morekeywords={WHERE,MATCH,RETURN,USING,PERIODIC,COMMIT,
      LOAD,CSV,WITH,HEADERS,FROM,CREATE,SET,WITH,SKIP,LIMIT,OPTIONAL,
      DISTINCT,EXISTS,AND},
    literate=
     *{0}{{{\color{numb}0}}}{1}
      {1}{{{\color{numb}1}}}{1}
      {2}{{{\color{numb}2}}}{1}
      {3}{{{\color{numb}3}}}{1}
      {4}{{{\color{numb}4}}}{1}
      {5}{{{\color{numb}5}}}{1}
      {6}{{{\color{numb}6}}}{1}
      {7}{{{\color{numb}7}}}{1}
      {8}{{{\color{numb}8}}}{1}
      {9}{{{\color{numb}9}}}{1}
      {:}{{{\color{punct}{:}}}}{1}
      {,}{{{\color{punct}{,}}}}{1}
      {\{}{{{\color{delim}{\{}}}}{1}
      {\}}{{{\color{delim}{\}}}}}{1}
      {[}{{{\color{delim}{[}}}}{1}
      {]}{{{\color{delim}{]}}}}{1}
}

\lstdefinelanguage{algorithm}{
    basicstyle=\scriptsize\ttfamily,
    numbers=none,
    numberstyle=\scriptsize,
    stepnumber=1,
    numbersep=8pt,
    showstringspaces=false,
    breaklines=true,
    frame=lines,
    backgroundcolor=\color{background},
    alsoletter={=, :},
    morekeywords={INPUT:, forEach, if, =, :, EntityType, StructuralVariation, RedisPair, RedisDatabase, String},
    literate=
      {,}{{{\color{punct}{,}}}}{1}
      {\{}{{{\color{delim}{\{}}}}{1}
      {\}}{{{\color{delim}{\}}}}}{1}
}

\lstdefinelanguage{json}{
    basicstyle=\scriptsize\ttfamily,
    numbers=none,
    numberstyle=\scriptsize,
    stepnumber=1,
    numbersep=8pt,
    showstringspaces=false,
    breaklines=true,
    frame=lines,
    backgroundcolor=\color{background},
    morekeywords={_id,_type,count,schema},
    literate=
     *{0}{{{\color{numb}0}}}{1}
      {1}{{{\color{numb}1}}}{1}
      {2}{{{\color{numb}2}}}{1}
      {3}{{{\color{numb}3}}}{1}
      {4}{{{\color{numb}4}}}{1}
      {5}{{{\color{numb}5}}}{1}
      {6}{{{\color{numb}6}}}{1}
      {7}{{{\color{numb}7}}}{1}
      {8}{{{\color{numb}8}}}{1}
      {9}{{{\color{numb}9}}}{1}
      {:}{{{\color{punct}{:}}}}{1}
      {,}{{{\color{punct}{,}}}}{1}
      {\{}{{{\color{delim}{\{}}}}{1}
      {\}}{{{\color{delim}{\}}}}}{1}
      {[}{{{\color{delim}{[}}}}{1}
      {]}{{{\color{delim}{]}}}}{1}
}

\title{A Unified Metamodel for NoSQL and Relational Databases\thanks{This
    work has been funded by the Spanish Ministry of Science, Innovation and
    Universities (project grant TIN2017-86853-P).}~$^,$\thanks{Formatted for
    arXiv.org.}}

\author{  \href{https://orcid.org/0000-0002-3835-9428}{\includegraphics[scale=0.06]{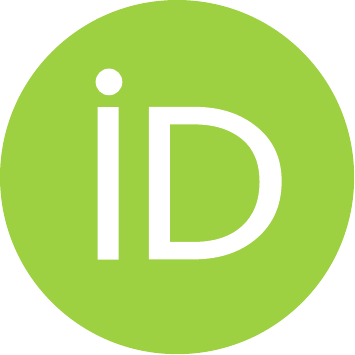}\hspace{1mm}Carlos J. Fernández Candel} \\
  Faculty of Computer Science\\
  University of Murcia\\
  Murcia, Spain\\
        \texttt{cjferna@um.es} \\
        \And \href{https://orcid.org/0000-0001-9313-008X}{\includegraphics[scale=0.06]{orcid.pdf}\hspace{1mm}Diego Sevilla Ruiz} \\
  Faculty of Computer Science\\
  University of Murcia\\
  Murcia, Spain\\
        \texttt{dsevilla@um.es}\\
        \And \href{https://orcid.org/0000-0003-4685-6659}{\includegraphics[scale=0.06]{orcid.pdf}\hspace{1mm}Jesús J. García-Molina} \\
  Faculty of Computer Science\\
  University of Murcia\\
  Murcia, Spain\\
        \texttt{jmolina@um.es}\\
}

\renewcommand{\shorttitle}{A Unified Metamodel for NoSQL and Relational Databases}

\begin{document}
\maketitle

\begin{abstract}

  The Database field is undergoing significant changes. Although
  relational systems are still predominant, the interest in NoSQL
  systems is continuously increasing. In this scenario, polyglot
  persistence is envisioned as the database architecture to be
  prevalent in the future. Therefore, database tools and systems are
  evolving to support several data models.

  Multi-model database tools normally use a generic or unified
  metamodel to represent schemas of the data model that they support.
  Such metamodels facilitate developing database utilities, as they
  can be built on a common representation. Also, the number of
  mappings required to migrate databases from a data model to another
  is reduced, and integrability is favored.

  In this paper, we present the \uschema{} unified metamodel able to
  represent logical schemas for the four most popular NoSQL paradigms
  (columnar, document, key-value, and graph) as well as relational
  schemas. We will formally define the mappings between \uschema{} and
  the data model defined for each database paradigm. How these
  mappings have been implemented and validated will be discussed, and
  some applications of \uschema{} will be shown.

  To achieve flexibility to respond to data changes, most of NoSQL
  systems are ``schema-on-write,'' and the declaration of schemas is
  not required. Such an absence of schema declaration makes
  \emph{structural variability} possible, i.e.,~stored data of the
  same entity type can have different structure. Moreover, data
  relationships supported by each data model are different; For
  example, document stores have \textit{aggregate objects} but not
  \textit{relationship types}, whereas graph stores offer the
  opposite. Through the paper, we will show how all these issues have
  been tackled in our approach.

  As far as we know, no proposal exists in the literature of a unified
  metamodel for relational and the NoSQL paradigms which describes how each
  individual data model is integrated and mapped. Our metamodel goes beyond
  the existing proposals by distinguishing entity types and relationship
  types, representing aggregation and reference relationships, and
  including the notion of structural variability. 
  Our contributions also include developing schema extraction strategies
  for schemaless systems of each NoSQL data model, and tackling performance
  and scalability in the implementation for each store.
\end{abstract}




\twocolumn

\section{Introduction\label{sec:introduction}}


With the advent of modern data-intensive applications (e.g.,~Big Data,
social networks, or IoT), NoSQL (\textit{N}ot \textit{o}nly \textit{SQL})
database systems emerged to overcome the limitations that relational
systems evidenced to support such applications, namely, scalability,
availability, flexibility, and ability to represent complex objects.
These systems are classified in several database paradigms, but commonly
the term NoSQL refers to four of them: \emph{columnar}, \emph{key-value},
\emph{document}, and \emph{graph} systems.
NoSQL systems of the same paradigm can have significant differences in
features and in the structure of the data. This is because there is no
specification, standard, or theory that establishes the data model of a
particular paradigm. Therefore, we will assume here the data model of the
most popular stores of each category. Table~\ref{table:TypesOfNoSQL} shows
the main features of the four mentioned NoSQL paradigms.

\begin{table*}[h!]
\centering
\begin{tabular}{@{}p{.1\textwidth}p{.25\textwidth}p{.4\textwidth}p{.2\textwidth}@{}}
  \toprule
  \multicolumn{4}{c}{{\bf NoSQL System Types}}\\
  \midrule
  {\bf Type}&{\bf Data structure}&{\bf Appropriateness} & {\bf Database Systems}\\
  \midrule
  Key-value&Associative array of key-value pairs& Frequent small read and writes
                                                  with simple data&Redis, Memcache\\
  Columnar&Tables of rows with varying columns. Column-based physical
            storage& High performance, availability, scalability, and large
                     volumes of data for OLAP queries&HBase, Cassandra\\
  Document&JSON-like document collections & Nested Objects, structural
                                            variation, and large volumes of
                                            heterogeneous data & MongoDB, Couchbase\\
  Graphs& Data connected in a graph& Highly connected objects, references
                                     prevail over nested objects & Neo4J, OrientDB \\
  \bottomrule
\end{tabular}
\caption{Types of NoSQL systems and some example
  implementations.\label{table:TypesOfNoSQL}}
\end{table*}


Over last years, as the popularity of NoSQL systems
increased,\footnote{Four of the top~10 being NoSQL systems in the DB
  engines ranking (\url{https://db-engines.com/en/ranking}) as of
  January~2021: MongoDB (5$^{\mathrm{th}}$), Redis (7$^{\mathrm{th}}$),
  Elasticsearch (8$^{\mathrm{th}}$), and Cassandra (10$^{\mathrm{th}}$).}
\emph{polyglot persistence\/} (a new term coined for heterogeneous database
systems) has been gaining acceptance as the data architecture of the
future: applications using the set of databases that better fit their
needs. Today, relational databases are still clearly the most used by a
wide margin, but most popular relational systems are evolving to support
NoSQL features.\footnote{Top-8 systems in the DB engines ranking
  (\url{https://db-engines.com/en/ranking} are multi-model.} Two facts have
motivated this interest in \emph{polyglot
  persistence\/}~\cite{stonebraker-blog-2015,fowler-nosql2012}:~(i)~the
complexity and variety of data to be managed by software systems,
and~(ii)~a single type of database system does not fit all the needs of an
increasing number of systems (e.g.,~learning management systems, online
retail systems, or social networks.)

The successful adoption of NoSQL requires database tools similar to those
available for relational systems. This entails to investigate how common
database utilities can be available for NoSQL systems. In addition, these
tools should be built taking into account the expected predominance of
polyglot persistence. Thus, they should support widespread relational
databases as well as NoSQL databases. In the case of data modeling tools,
the shift towards multi-model solutions is evident: the most popular
relational modeling tools are being extended to integrate NoSQL stores
(e.g.~ErWin\footnote{\url{http://erwin.com/products/erwin-data-modeler}.}
and
ER/Studio\footnote{\url{https://www.idera.com/er-studio-enterprise-architecture-solutions}.}
provide functionality for MongoDB) and new tools supporting a number of
relational and NoSQL databases have appeared
(e.g.~Hackolade\footnote{\url{https://hackolade.com/}.}). This multi-model
nature should be considered for database tooling, i.e.,~tools should
support multiple data models.

\emph{Data models} determine how data can be organized and manipulated in
databases. They are applied to a particular domain by defining
\emph{schemas} that express the structure and constraints for the domain
entities. Such information, provided by schemas, is necessary to implement
many database tools. However, most NoSQL systems are \emph{schemaless}
(a.k.a ``schema-on-read''), that is, data can be directly stored without
requiring the previous declaration of a schema. This feature is motivated
by the fact that the pace of data structure changes is considerably faster
in the new data-intensive applications.
Being \emph{schemaless} does not mean the absence of a schema, but that the
schema information is implicit in data and code. Therefore, the schemas
implicit in NoSQL stores must be reverse engineered in order to build a
cohesive set of utilities for NoSQL databases such as schema viewers, query
optimizers, or code generators. This reverse engineering process must
tackle the fact that ``schema-on-read'' systems can store data with
different structure even belonging to the same database entity type (and
relationship type in graphs), i.e.,~each entity or relationship type can
have one or more \emph{structural variations\/}.
Recently, several NoSQL schema extraction approaches have
been published~\cite{sevilla-er2015,klettke-schema2015,wang-schema2015},
and some data modeling tools such as those mentioned above provide some
kind of reverse engineering functionality.

When building multi-model database tools, the definition of a generic,
universal, or unified metamodel can provide some
benefits~\cite{englebert1999db,bernstein2007b,atzeni2009,kensche2007,allenwang2016}.
It offers a unified view of different data models, so that their schemas
will be represented in a uniform way. This uniformity facilitates building
generic tools that are database-independent. With the predominance of
relational databases, the interest in multi-model tools declined, and
little attention has been paid to the definition of unified metamodels. A
remarkable proposal is the DB-Main approach~\cite{englebert1999db,hick2003}
that defined the GER generic metamodel based on the EER (Extended Entity
Relationship) data model. More recently, some universal
metamodels~\cite{atzeni2009,kensche2007} have been created in the context
of ``Model Management''~\cite{bernstein2007b,bernstein2000}. With the
emergence of NoSQL systems, some unified metamodels have been proposed to
have a uniform access to data~\cite{atzeni2009,atzeni2012}, and the idea of
an unified metamodel for data modeling tools was outlined
in~\cite{allenwang2016}.

In the past years, we defined a reverse engineering strategy to infer
logical schemas from document NoSQL databases~\cite{sevilla-er2015}, and
have presented approaches to visualize inferred
schemas~\cite{alberto-erforum2017} and automatically generate
object-document mappers code~\cite{alberto-mappers2019} from schemas.
Currently, we are tackling the definition of a synthetic data generation
approach~\cite{alberto-comonos2020}, and the extraction of implicit
physical schemas from NoSQL data~\cite{pablo-comonos2020}. As our intention
is to build multi-model database utilities, we have tackled the definition
of a unified metamodel named \uschema{}, which is capable of representing
schemas for the four most common NoSQL data models, as well as the
relational model. In this paper, we present \uschema{} jointly with a data
model for each NoSQL paradigm, and the mapping from those data models to
\uschema{} (\emph{forward mappings\/}), and from \uschema{} to the data
models (\emph{reverse mappings\/}). Common strategies are defined to
implement and validate the mappings.
Several applications of \uschema{} are then commented.

The \textbf{research contributions} of this work are as follows:

\begin{enumerate}
\item To our knowledge, we present the first unified metamodel able to
  represent logical schemas both for the four most common kinds of NoSQL
  systems and relational systems. Two salient features of our proposal
  are:~(i)~\uschema{} includes the notion of \emph{structural variation}
  for entity and relationship types, as most NoSQL systems are schemaless,
  and~(ii)~unlike other proposals, the four kinds of relationships between
  entities that are typical in logical data modeling are supported by
  \uschema{}: aggregation, references, graph relationships, and
  generalization. Capturing structural variability allows us to accurately
  describe the structure of the stored data.

\item Defining the forward and reverse mappings between \uschema{} and the
  data models that it integrates, we have established the notion of
  \emph{canonical mapping} in which there is a natural correspondence
  between each element of a data model and elements of \uschema{}. For each
  data model, its canonical mapping has been formally defined, as well as
  the reverse mapping for characteristics not present in the considered
  data model (e.g.~graph systems do not support aggregate relationships, or
  structural variation is not possible in relational tables.)


\item A common strategy is proposed to extract unified schemas from
  databases. This strategy has been applied to implement canonical mappings
  for each paradigm integrated in \uschema{}. As far as we know, all the
  schema extraction proposals for NoSQL stores, only consider one kind of
  store, normally document-based~\cite{klettke-schema2015,wang-schema2015}
  or graph~\cite{comyn-wattiau2017}. Our strategy also takes into account
  scalability and performance, using MapReduce processing on the native
  data.

\item We provide insights on how \uschema{} can ease the implementation of
  database utilities in multi-model and multi-database environments such as
  database migration, schema queries, data generation for testing, query
  optimization, and schema visualization.


\item We have defined the \uschema{} metamodel with the Ecore metamodel,
  which is the central element of the Eclipse Modeling Framework
  (EMF)~\cite{steinberg-emf2009}, a widely used open-source platform to
  develop \emph{Model-Driven Software Engineering\/} (MDE)
  solutions~\cite{brambilla2012}. Thus, the proposal is open to be used and
  incorporated in any future database tool development.
\end{enumerate}

The rest of this paper is organized as follows. Next section will present
the \uschema{} data model. Then, Section~\ref{sec:commonstrategy} will
describe the common process devised to reverse engineer implicit schemas
from data, introduce a running example database, and will explain the
common strategy applied to validate and assert the performance of the
schema extraction algorithms. Next five sections will be devoted to define
a logical data model, formally specify the mapping between \uschema{} and
the data model defined, and show the implementation and validation of the
corresponding forward mapping. Once presented the unified metamodel and the
mappings, we will discuss how they can be applied in common database tasks.
Finally, we will contrast our proposal with related works, draw some
conclusions, and outline further work.

\section{The \uschema{} Unified Data Model\label{sec:uschema-model}}

\subsection{Logical Modeling Concepts in
  \uschema{}\label{sec:uschema-concepts}}

A data model provides a set of concepts to specify the structure and
constraints of a database type, and a \emph{schema} results of applying a
concrete data model on a domain or problem. A schema is therefore an
instance of a data model. Given a particular data model, textual and
graphical languages can be defined to express schemas.

Data models (and therefore schemas) can be defined at different levels of
abstraction. Typically, they are classified in three categories: conceptual,
logical, and physical. \emph{Conceptual} schemas represent the domain of an
application in a platform-independent way. \emph{Logical} schemas describe
data structures and constraints, but providing physical independence.
Finally, \emph{Physical} schemas include all details needed to implement a
logical schema on a specific database system.

At the logical or physical level, a \emph{unified} or \emph{generic data
  model} can be defined to integrate concepts from several data models with
the purpose of offering a uniform representation. When using a unified
model for $n$ data models, instead of managing $n \times (n-1)$ mappings
(each data model with the others), only $n+n$ mappings are needed (between
the unified and each of the integrated data models in both directions.)

\uschema{} is a unified logical model that integrates the concepts and
rules of both the relational model and the four most common NoSQL data
models: columnar, document, graph, and key-value. While the relational
model is a well-defined data model, there is no specification, standard, or
theory that establishes the data model of a particular NoSQL paradigm. In
fact, NoSQL systems of the same kind can have significant differences in
features and in the structure of the data. We have therefore defined a
logical model for each NoSQL category by abstracting from the
logical/physical data organization of the most popular stores of each
category. This section will present \uschema{}, while the logical model
defined for each particular NoSQL paradigm will be presented in the section
devoted to describe how that data model has been integrated into
\uschema{}.


\uschema{} includes the basic concepts traditionally used to create logical
schemas, which are part of well-known formalisms such as
\emph{Entity-Relationship} (ER)~\cite{chen-1976} and \emph{UML Class
  Models}~\cite{rumbaugh-uml1999}: entity type, simple and multivalued
attributes, key attribute, and three kinds of relationships between entity
types: aggregation, reference, and inheritance. Also, \uschema{}
incorporates some additional concepts, such as \emph{relationship types}
(as they are considered in the graph data model~\cite{angles2017}), and
\emph{structural variations} of entity and relationship types. Before
presenting the \uschema{} metamodel, we will define all these concepts. Not
all concepts will be present in all of the data models supported by
\uschema{}. For example, the relationship type is exclusive of the graph
model, but conversely, it does not define aggregation.

In data models, an entity type $\varepsilon$ is normally characterized by a
set $P^\varepsilon=\{P^{\varepsilon}_i\}, i=1 \ldots n$ of named
properties. Properties can be of several kinds depending on the type of the
object or value a property can hold. Three common kinds are: attributes,
aggregations, and references. Given a property $P^{\varepsilon}_i$, it
would be an attribute if it can take values of scalar type (e.g.~Number) or
structured type (i.e.~Array or Set), and it would be an aggregation or
reference if it is associated to an entity type ${\varepsilon}'$ whose
objects are, respectively, embedded in or referenced from objects of the
entity type $\varepsilon$, to which the $P^{\varepsilon}_i$ property
belongs. Keys are a special kind of attribute able to record 
values used as identifiers of instances of entity types.

Graph data models include \emph{relationship types} in addition to entity
types. While nodes are instances of entity types, arcs are instances of
relationship types, which can have attributes. Hereafter, we will use
``schema type'' to gather both entity and relationship types.


Schemas play a similar role to types in programming languages. Given a
database schema $S$, only data conforming to $S$ can be stored in the
database. Therefore, all data of an entity type $E$ (resp. a relationship
type $R$) will have the same structure, that defined for $E$ (resp.~$R$) in
$S$. In absence of schema declarations, however, data of $E$ and $R$ can
have different structure, that is, $E$ and $R$ will have one or more
\emph{structural variations}.


A structural variation of a schema type $\varepsilon$ is formed by a set of
properties $Q^\varepsilon \subset P^\varepsilon$, and each pair of
variations of $\varepsilon$ differ, at least, in one property. The set
$P^\varepsilon$ is therefore the union of the sets of properties of each of
its variations. The set $P^\varepsilon$ is commonly called \emph{union
  schema} of a schema type in a schemaless system. The properties of an
schema type can therefore be classified as \emph{common} or
\emph{specific}, depending on whether they are present in all the
variations, or in a subset of them. It is worth noting that specifying a
schema type as the \emph{union schema}, the information on its structural
variability is lost. We have considered convenient to record this
structural variability in data models defined for NoSQL databases, and
therefore the notion of ``variation'' will be part of \uschema{}.

When structural variations are considered, entity and relationship types can
be defined as follows.


\begin{itemize}
\item An \textbf{entity type} has a name and is formed by a set of
  structural variations.


\item A \textbf{relationship type} (only for graph stores) has a name, is
  formed by a set of structural variations, and refers to both a source and
  a target entity type.

\item A \textbf{structural variation} is formed by a set of named
  properties. The kind of properties depend on the data model, and can be:
  attributes, keys, aggregates, and references.
\end{itemize}

Table~\ref{table:uschema-mappings} shows the correspondence between
concepts of each of the considered database kinds and the logical modeling
concepts that we will use in \uschema{}. We will use these concepts to
abstract a \emph{logical data model} for the most popular systems of each
NoSQL paradigm. These models will be presented in
sections~\ref{sec:graph-database}~to~\ref{sec:columnar-inference}.

\begin{table*}[ht!]\centering
\scalebox{.92}[.92]{%
\noindent%
\begin{tabular}{p{.2\textwidth}p{.15\textwidth}p{.15\textwidth}p{.15\textwidth}p{.15\textwidth}p{.15\textwidth}}
\toprule
\raggedright\textbf{Logical modeling concepts} & \textbf{Relational} & \textbf{Columnar} & \textbf{Document} & \textbf{Graph} & \textbf{Key/Value} \\
\midrule

\textit{Schema} & \raggedright Schema & \raggedright Database or Keyspace & Database & Graph &  \raggedright Database or Namespace \tabularnewline
\addlinespace[0.5em]

\textit{Entity Type} & Table & \raggedright Table with column families & \raggedright Collection and nested object & \raggedright Node label & \raggedright Multirow entities \tabularnewline
\addlinespace[0.5em]

\textit{Relationship Type} & Relationship Table & N/A & N/A & Relation type & N/A \\
\addlinespace[0.5em]

\raggedright \textit{Structural Variation} & \raggedright Table (only one variation) & \raggedright Rows with different structure within column families & \raggedright Documents with different structure in a collection & \raggedright Same label with different structure & \raggedright Multirow entities with different structure \tabularnewline
\addlinespace[0.5em]

\textit{Key} & Primary key & Row key & Document key & N/A & Pair key \\
\addlinespace[0.5em]

\textit{Reference} & Foreign key & \raggedright Join between tables & \raggedright Join between documents & N/A & \raggedright Join between pairs \tabularnewline
\addlinespace[0.5em]

\textit{Aggregation} & N/A & \raggedright Nested object & \raggedright Nested object & N/A & \raggedright Nested object  \tabularnewline

\addlinespace[0.5em]

\textit{Attribute} & Column & Column & \raggedright Document property & \raggedright Node and Relation property & Pair Value \\
\addlinespace[0.5em]

\textit{Primitive Types} & Scalar Types & Scalar Types & Scalar Types & Scalar Types & Scalar Types \\
\addlinespace[0.5em]

\textit{Structured Types} & N/A & Collections & Collections & Array & Collections \\
\bottomrule
\end{tabular}%
}%
\caption{Mapping between logical modeling concepts and NoSQL/Relational
  Database Systems.\label{table:uschema-mappings}}
\end{table*}

\subsection{The \uschema{} Metamodel\label{sec:uschema-metamodel}}

\begin{figure*}[!htb]
  \includegraphics[width=\textwidth]{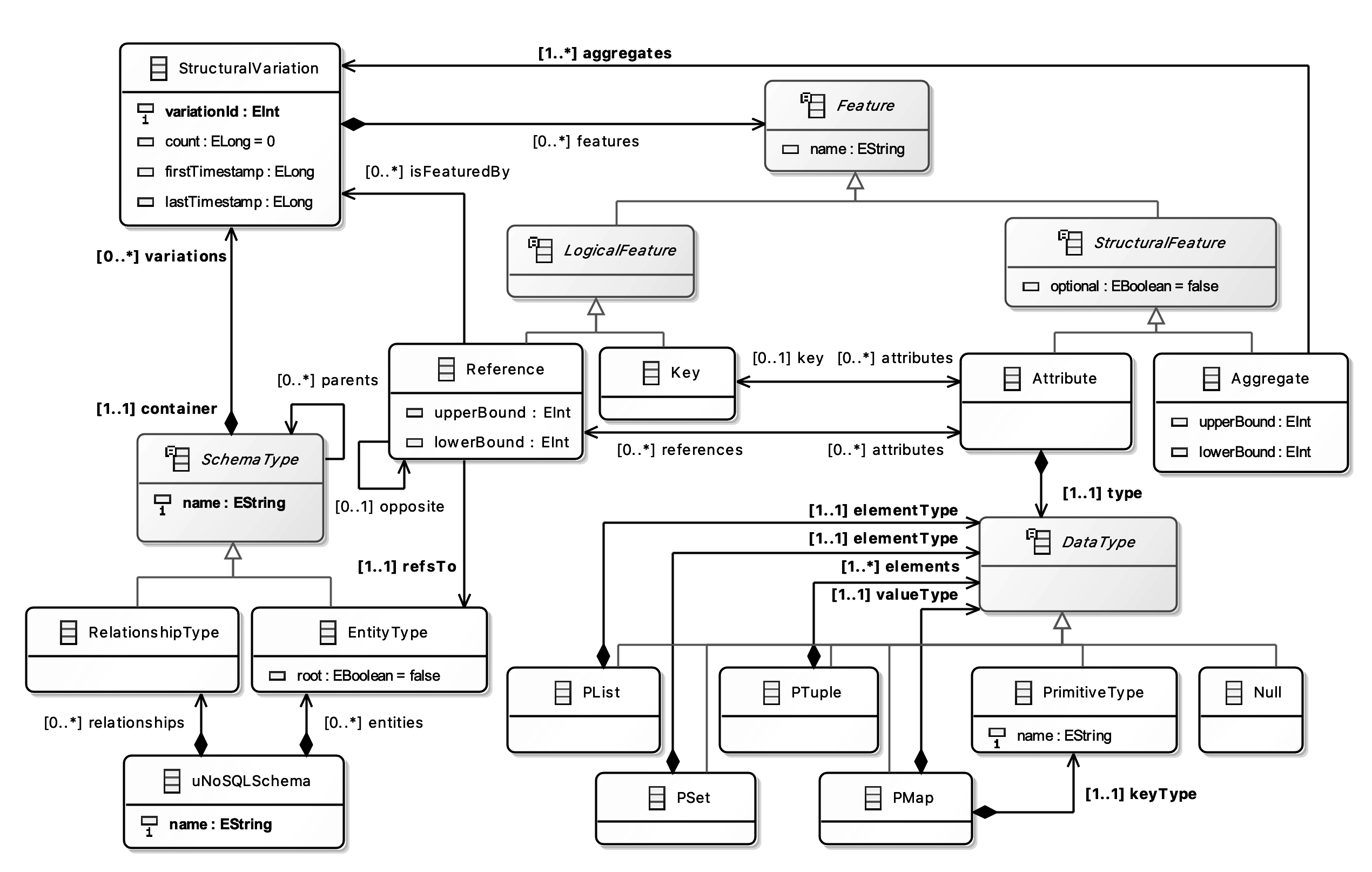}
  \caption{\uschema{} Metamodel.\label{fig:uschemametamodel}}
\end{figure*}

Data models are commonly expressed formally in form of \emph{metamodels}. A
metamodel is a model that describes a set of concepts and relationships
between them. That description determines the structure of models that can
be instantiated from the metamodel elements, i.e.~\emph{a metamodel is a
  model of a model\/}. Object-oriented conceptual modeling is usually
applied to create metamodels: concepts and their properties are modeled
with classes, and reference, aggregation, and inheritance relationships are
used to model relationships between concepts.
Figure~\ref{fig:uschemametamodel} shows the metamodel of the \uschema{}
data model in form of a UML class diagram. Below, we describe this
metamodel.

A \texttt{\uschema{}} model represents a schema formed by a collection of
types (\texttt{SchemaType}) that can be either entity types
(\texttt{EntityType}) or relationship types (\texttt{RelationshipType}).
Both types have two common properties: They include one or more structural
variations (\texttt{StructuralVariation}), and they can form a type
hierarchy (\texttt{parent} relationship).

A \texttt{StructuralVariation} has an \texttt{identifier} and is
characterized by a set of \emph{logical} and \emph{structural} features.
\texttt{StructuralFeature}s determine the structure of database objects,
and include \texttt{Attribute}s and \texttt{Aggregate}s, while logical
features specify what identifies an object (\texttt{Key}), and which
\texttt{Reference}s an object has to other objects.

Each attribute has a name and a data type. The data types included are:
\texttt{Primitive} (e.g.,~Number, String, Boolean), \emph{collections}
(sets, maps, lists, and tuples), and the special \texttt{Null} type. Also,
the JSON and BLOB primitive types are included to support relational
systems. An aggregation has a name, a cardinality (upper and lower bound),
and refers to the structural variation it aggregates, or to a list of
variations, if the aggregated object is an heterogeneous collection.

Unlike aggregations, references refer to an entity type (via
\texttt{refsTo}), and one or more \texttt{attributes} that match the set of
key attributes of the referenced object (all the variations of an entity
type must have the same key.) References also have a name, a cardinality,
and an optional inverse reference (\texttt{opposite}). Additionally,
references can have their own attributes when they represent graph arcs.
This entails that a reference has to specify which variation (of its
\texttt{RelationshipType}) its set of attributes corresponds to
(\texttt{isFeaturedBy}). \texttt{Key} represents the set of attributes
playing the role of key for an entity type, holding a unique set of values
for each element of the type. \texttt{Reference} also points to the set of
attributes that form the referenced key (\texttt{attributes} property).

The aggregation relationship allows objects to be recursively embedded,
then forming aggregation hierarchies. In these hierarchies, the
identification of the root element is important. Thus, an entity type
includes a boolean attribute named \texttt{root} to indicate whether or not
their entities are aggregated by others (\texttt{aggregates} relationship).

\uschema{} also records information that could be useful to implement some
database tasks. For example, as shown in Figure~\ref{fig:uschemametamodel},
\texttt{StructuralVariation}s have a \texttt{count} attribute to record the
number of objects that belong to each variation, and two timestamps that
hold the creation dates for the first and last stored object of a variation
(\texttt{firstTimestamp} and \texttt{lastTimestamp}).



The \uschema{} metamodel has been defined with the Ecore metamodeling
language. Ecore is the central element of \emph{Eclipse Modeling Framework}
(EMF)~\cite{steinberg-emf2009}, a framework widely used to develop
Model-Driven Software Engineering (MDE) solutions~\cite{brambilla2012}.
EMF-provided tools such as model transformation languages, model comparison
and diff/merge tools, or workbenches for the creation of domain-specific
languages (DSLs) could be used to build database tools based on \uschema{}
models. Metamodeling has traditionally been applied to define data models,
and transformational approaches have been proposed to tackle problems
involving schema mappings~\cite{hainaut2006,bernstein2000}. However, the
database engineering community has paid little attention to MDE techniques
and tools, although metamodeling and model transformations foundations are
well established in the MDE field. Using Ecore, we obtain two benefits:
leveraging the EMF tooling to develop database utilities, and favor their
interoperability with other tools~\cite{bermudez2017}.

\subsection{\uschema{} Flavors: Full Variability vs. Union Schema}

\uschema{} allows to accommodate the definition of two model flavors:

\begin{itemize}
\item {\bf Full Variability}: All structural variations of all entity and
  relationship types are stored.
\item {\bf Union Schema}: Only one structural variation is stored for
  each schema type. Structural variability is recorded by using the
  \texttt{optional} boolean attribute of {\tt Feature} to indicate if a
  feature is present or not in all the objects of an schema type. Union
  schemas are the schemas normally obtained in NoSQL schema discovering
  processes~\cite{klettke-schema2015,wang-schema2015}, and visualized in
  NoSQL modeling tools.
\end{itemize}

Note that it is easy to convert a \uschema{} model from the \emph{Full} to
the \emph{Union} flavors. This conversion loses information, and thus it is
not reversible: Given a schema type $t$ with a set of $n$ variations
$t.variations = \{V^{t}_i\}, i=1 \ldots n,$, then $t$ will be replaced by a
schema type $s$ with the same name ($t.name = s.name$), and the set
$s.variations$ will have a single variation $W^s$ with
$W^s.features = \biguplus_{i=1}^n V^{t}_i$, where $\biguplus$ is a function
that returns the union set of all the features of all the variations with
the following rules:

\begin{enumerate}
\item If the same structural feature appears in all variations $V^t_i$,
  then it is included in the result set with its {\tt optional} attribute
  set to false (common structural features).
\item Each structural feature that appears at least in a variation is
  included in the result set, but with its {\tt optional} attribute set to
  true.
\item Structural features that appear with the same name ({\tt name}
  attribute of {\tt StructuralFeature}) but with different type (they
  belong to a different sub-metatype of {\tt StructuralFeature} or have
  different values of their attributes), are included with a numeric
  identifier appended to their {\tt name}, and with their {\tt optional}
  attribute set to true.
\end{enumerate}

Example of an union schema for the running example presented in
Section~\ref{running-example} is shown in
Figure~\ref{fig:union_schema_graph}. {\tt StructuralVariation}s are omitted
for clarity, and optional features are shown in \emph{cursive} and green
color.

We will use the Full Variability flavor through the rest of the article, as
it contains more information and can be trivially converted to the Union
Schema if desired.

\subsection{Mappings between \uschema{} and the Logical Data
  Models\label{sec:uschema-mapping}}

A unified metamodel is intended to represent all the concepts of the
individual data models that it integrates. Therefore, a mapping must be
established between the unified metamodel and each data model. We will call
\emph{forward mapping\/} to a mapping from a NoSQL or relational model to
\uschema{}, and \emph{reverse mapping\/} to a mapping in the opposite
direction.

As indicated above, we had to define a logical data model for each NoSQL
paradigm. As most NoSQL databases are schemaless, the schemas are implicit
in data and code. Therefore, the implementation of a forward mapping must
first capture all the logical information of the implicit schema, as
described in Section~\ref{sec:commonstrategy}, and then apply the mapping
to obtain the \uschema{} schema (i.e.,~a \uschema{} model).

As \uschema{} is richer in concepts than each individual data model,
forward and reverse mappings are not unique for a particular data model.
In addition, \uschema{} concepts not present in a specific data model
could be mapped in different ways in a reverse mapping. This has led us to
introduce the notion of \emph{canonical mapping}. A canonical mapping
satisfies two conditions:

\begin{enumerate}
\item It must be \emph{forward-complete}, that is, the rules must correctly
  map all the characteristics of the data model to \uschema{} concepts.
\item As a consequence, it must be trivially \emph{bidirectional within a
    data model}. This is because given a \uschema{} model, the original
  database schema could always be reproduced (as the \uschema{} model holds
  all its information.)
\end{enumerate}

While the canonical mapping rules cover the characteristics of each of the
logical data models, there may be cases where a reverse mapping has to be
performed on a \uschema{} model that contains elements not present in a
given data model.
Specialized forward and reverse mappings could also be defined for each
data model, and even for a given database implementation. These mappings
could be devised for specific needs within a development such as a database
migration that involves different source and target data models. The need
for these mappings raises the interest in creating a mapping language able
to specify how the constructs of a given database paradigm are translated
into the abstractions of \uschema{}, and vice versa. This is out of the
scope of this work.

In the following Section, the common strategies devised to implement and
validate all the forward mappings will be described. In
sections~\ref{sec:graph-database} to~\ref{sec:relational-model}, we will
define a logical model for each database paradigm addressed and formally
express the \emph{canonical mapping\/} between each data model and
\uschema{}. Additionally, reverse mapping examples will be shown for
characteristics not supported in each of the data models. For each
paradigm, the forward mapping implementation and validation will be
commented. Here, we will introduce the notation used to define
the mappings.




\begin{itemize}
\item A mapping between an element $u$ of \uschema{} and an element $m$ of
  a data model is expressed as:
  \[u \leftrightarrow m \| \textnormal{\{list of \emph{property
        relations}\}}\]
  where \emph{property relations} are expressed as indicated below, and the
  $\leftrightarrow$ operator is commutative.

\item A property relation $p_1 = p_2$ expresses that a property $p_1$ of
  $u$ and an property $p_2$ of $m$ have the same value. The $=$ operator is
  commutative.

\item A property relation $p \leftarrow v$ expresses that the value $v$ is
  assigned to the $p$ property of $u$ or $m$.

\item Let $e_1$ be a property of $u$ and let $e_2$ be a property of $m$, a
  property relation $e_1 \leftrightarrow e_2$ expresses an enclosed mapping
  between $e_1$ and $e_2$.


\item In a property relation that expresses a mapping between two elements,
  the $map(e,t)$ function can be used to obtain the target element of type
  $t$ that maps to the source element $e$; if $e$ maps to a single target
  element, then the second argument is optional.

\item Given an instance of a meta-class in \uschema{}, dot notation is used
  to refer to its properties.
  For example, given an instance $e$ of \texttt{EntityType}, $e.name$
  refers to the attribute \texttt{name}.

\item Functions will be applied on elements of the data model to obtain the
  value of their properties. Functions will have the same name as the
  property. For example, given an entity type $e$, $name(e)$ will refer to
  its \textit{name} property. Other functions will be introduced in some
  rules, and their proper definition will be shown.
\end{itemize}



\section{A Common Strategy for the Implementation and Validation of the
  Extraction of \uschema{} Models\label{sec:commonstrategy}}



In this Section, we will first explain how \uschema{} models are built from
NoSQL databases or relational schemas. Then, a conceptual schema will be
presented as a running example to be used to illustrate the explanations of
the following five sections. Finally, the experiments used to validate the
\uschema{} model building process will be exposed. The implementation and
validation strategies are common for all the paradigms, but some stages or
experiments are not required in the case of the relational model.

\subsection{Building \uschema{} Models\label{common-implementation}}

\begin{figure}[!ht]
  \centering
  \includegraphics[width=0.5\textwidth]{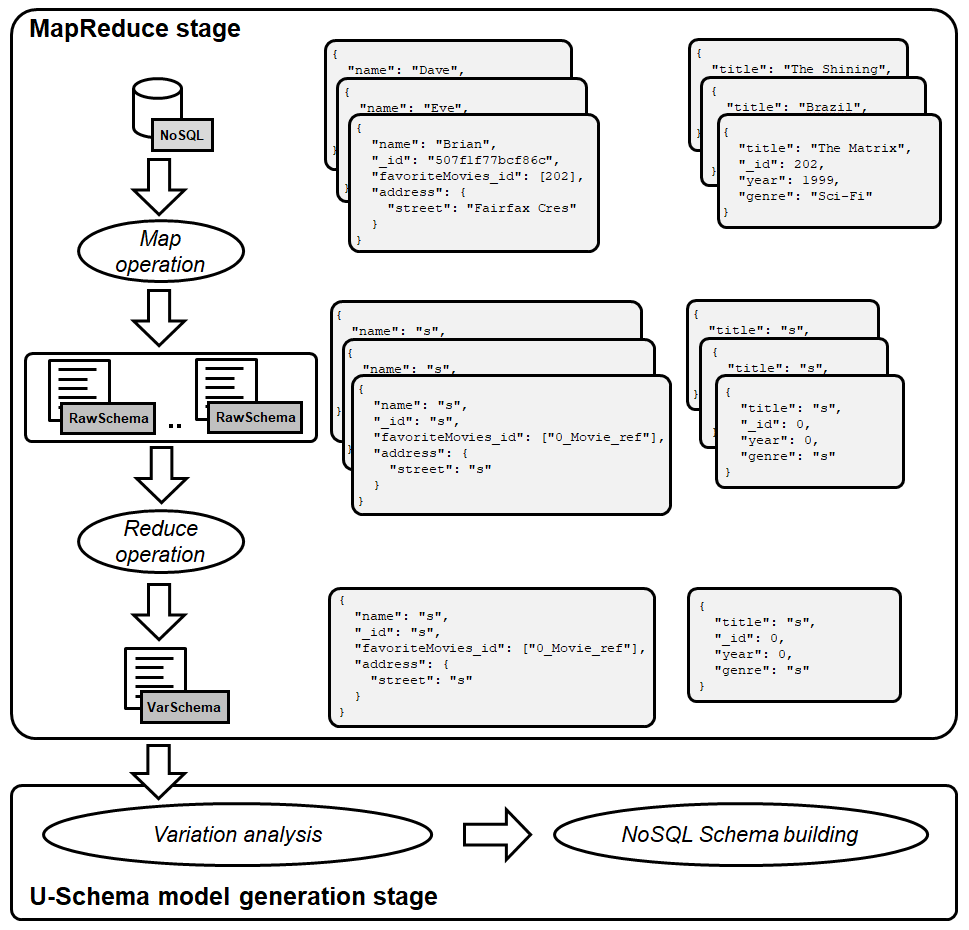}
  \caption{Generic Schema Extraction Strategy.\label{fig:phases}}
\end{figure}

As indicated in Section~\ref{sec:uschema-mapping}, in the case of NoSQL
stores, applying a forward mapping first requires inferring the schema that
is implicit in the data and code. These schemas conform to the
\emph{logical data model} abstracted for each NoSQL paradigm. Therefore,
\uschema{} models are built in a~2-stage process, as illustrated in
Figure~\ref{fig:phases}. First, a MapReduce operation is performed on the
database to infer its logical schema. This stage is not needed for the
relational model.
In the second stage, the forward mapping rules are applied to create a
\uschema{} model from the previously inferred schema. Next, we explain
these two stages.

\paragraph{Inferring the Logical Schema}



In the \emph{map} operation, a \emph{raw schema} is obtained for each
object stored in the database. We call \emph{raw schema} to an intermediate
representation (JSON-like format) that describes the \emph{data structure
of a structural variation}: a set of pairs formed by the name of a
property and its data type. Given an object $O$ stored in the database of
an entity type $e$, its raw schema is obtained by applying the
following~4~rules on the values of its properties $p_i$:

\begin{enumerate}
\item [R1] Each value $v_i$ of a $p_i$ property is replaced by a value
  representing its type according to the rules R2 and R3.
\item [R2] If $v_i$ is of scalar or primitive type, it is replaced by a
  value that denotes the primitive type: {\tt "s"} for
  String,~\texttt{0}~for numeric types, \texttt{true} for Boolean, and so
  on.
\item [R3] If $v_i$ is an embedded object, the rules R1, R2, and R3 are
  recursively applied on it.
\item [R4] If $v_i$ is an array of values or objects, rules R2 and R3 are
  applied to every element, and the array is replaced with an array of
  values representing types.
\end{enumerate}

In the case of document systems, where the key is explicitly included in
the documents, the representation of the structure will contain one scalar
property with the name ``{\tt \_id}'', representing the key of the entity
type. Additionally, the following rule is applied to infer references
between objects:

\begin{enumerate}

\item[R5] Some commonly used conventions and heuristics are taken into
  account to identify references. For example, if a property name (with an
  optional prefix or suffix) matches the name of an existing entity type
  and the property values match the values of the ``{\tt \_id}'' property
  of such an entity. The value of the property is replaced concatenating
  the value indicated in rule R2 with the name of the entity type and the
  suffix ``{\tt \_ref}.''
\end{enumerate}

The process is repeated to obtain the \emph{raw schemas} of the
relationship types in the case of graph databases.


Figure~\ref{fig:phases} shows how the above rules are applied to
\emph{User} and \emph{Movie} objects of a document store. A raw schema is
obtained for each \emph{User} object with identical structure, and the same
for \emph{Movie} objects.


Once the map function is performed, the reduce function collects all the
identical raw schemas and outputs a single representative raw schema for
each structural variation of an entity type, to which we will refer,
hereafter, as \emph{variation schema}. Figure~\ref{fig:phases} shows the
variation schemas obtained for \emph{User} and \emph{Movie} objects. Note
that a \emph{variation schema} will be generated for each structural
variation of the objects.

In the case of graph and key-value systems, a preliminary stage is needed
to achieve an efficient MapReduce processing, as explained in
Sections~\ref{sec:graph-mapping-implementation}
and~\ref{sec:redis-mapping-implementation}.

We decided to build \uschema{} models directly from the intermediate
representation of the MapReduce output instead of building specific
metamodels for each paradigm, because \uschema{} already contains the
abstractions present in each of the individual data models, and the
transformation would have been redundant.

\paragraph{Generating a \uschema{} Model}

In the second stage, \emph{variation schemas} are analyzed to build the
\uschema{} model. For this, a parsing process is connected to a schema
construction process by applying the Builder pattern~\cite{gamma1994}.
Variation schemas are parsed to identify its constituent parts: properties
and relationships, as well as the entity type (or relationship type) to
which they belong. Whenever the parser recognizes a part, it passes it to a
builder that is in charge of creating the schema. A builder has been
implemented for each data model, which captures how parts are mapped to
\uschema{}. The same parser is used for all the data models as its input
are variation schemas. In the case of relational databases, only this
second stage is needed, as schemas are already declared.

\subsection{The ``User Profiles'' Running Example\label{running-example}}

\begin{figure}[!ht]
  \centering
  \includegraphics[width=\linewidth]{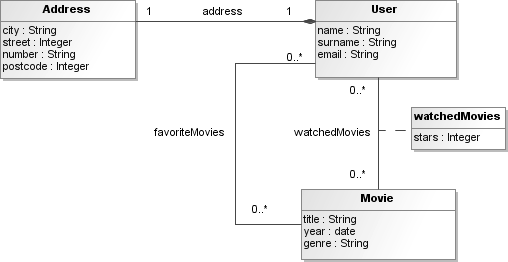}
  \caption{``User profile'' running example
    schema.\label{fig:running-example-schema}}
\end{figure}

Figure~\ref{fig:running-example-schema} shows a ``User Profiles''
conceptual schema that will be used to build a database example for each
paradigm integrated in our unified model. In each case, a ``User Profiles''
database will be populated to execute the algorithm that creates the
corresponding \uschema{} model and also to validate this algorithm as
explained in the next subsection.

``User Profiles'' schema could be an excerpt of the conceptual schema of a
movie streaming platform, which is expressed as a UML class model. It
has~3~entities labeled \textit{Movie}, \textit{User}, and \textit{Address},
and~3~relationships: a user aggregates an \emph{address}, a user has zero
or more \emph{favorite movies}, and a user has zero or more \emph{watched
  movies}. \textit{User} has the attributes \textit{name},
\textit{surname}, and \textit{email}; \textit{Address} has \textit{city},
\textit{street}, \textit{number}, and \textit{postcode}; and \textit{Movie}
has \textit{title}, \textit{year}, and \textit{genre}.

When instantiating each database, we will suppose that there
are~2~variations for the \textit{Address} entity type:
$\{street, number, city\}$, and $\{street, number, city, postcode\}$;
and~2~variations for \textit{User} that vary in the relationships: either
\textit{favoriteMovies} and \textit{watchedMovies} coexist, or only
\textit{watchedMovies} is present, and in the attributes: the
\textit{surName} attribute is only present when both relationships are.

\subsection{Validation of the Schema Building
  Process\label{common-validation}}

To validate our schema building process, we have applied the
same validation for the four kinds of NoSQL paradigms. For each system, we
used two databases, a synthetic one based on the running example, an a real
dataset. In each one of them, two experiments were carried out:~(i)~a
round-trip strategy to check that the obtained \uschema{} model is
equivalent to the schema used to synthesize the database or the schema of
the real existing database; and~(ii) two queries are issued on the real and
synthesized databases to assure that at least a data object exists for each
inferred structural variation (``all variations exist'' query) and that the
extraction process correctly calculates the number of data objects of each
variation (``correctness count'' query). In the case of the relational
model, only the second experiment is performed, as only the canonical
forward mapping must be implemented, because there is no need to infer the
logical model of the database.

The round-trip experiment consisted in the following steps. First, we
manually created a \uschema{} model (i.e.~a schema) with the desired
database structure (or the existing structure in the case of the real
database). The running example model covers all the elements that can be
mapped into the logical data model of the corresponding paradigm, but this
may not be the case for the real dataset. To populate the initial running
example database, we randomly created elements according to the defined
model. Afterwards, we inferred the implicit schema, and finally verified
that this schema was equivalent to the original \uschema{} model: the
resulting model can differ in the ordering of the different variations
found for each entity or relationship type, this is why in this case we
could not use standard model comparison tools, so we built a custom
\uschema{} model compare utility.

To evaluate the scalability and performance of the \uschema{} model building
algorithms, we have generated four datasets of different size for the
running example. The larger database contains~\numprint{800000} objects for
the \emph{User} and \emph{Address} entities,~\numprint{400000} for
\emph{Movie}, and a mean of~20 watched movies and~20 favorite movies.
\emph{User} and \emph{Address} have the same number of objects in each of
their variations. The rest of datasets reduce the number of objects and
relationships in a factor of~2,~4, and~8, as shown in
Table~\ref{tab:database-sizes}.

All the performance tests were run on an Intel(R) Core(TM) i7-6700 CPU
@~3.40GHz with~48~GB of RAM and using SSD storage. To give a meaningful
expression of the scalability of the schema inference process, instead of
comparing absolute times, we used as a time baseline an aggregate query
that calculates the average of watched movies by users. This query could be
representative of those obtaining periodic reports, so we suppose that the
database is not optimized for it. In this way, we can get results that are
independent of the different configurations in the deployment.
Table~\ref{tab:database-times} show the different times for the queries,
schema inference, and the normalized value (inference time divided by query
time) for the database sizes in Table~\ref{tab:database-sizes}. We expected
the ratio to diminish as the size of the database increases, as the
initialization time of the MapReduce framework becomes smaller with respect
to the total inference time. Moreover, in all cases the ratio stays in the
range of \numprint{17.58}x (MongoDB, smaller case) to \numprint{2.04}x
(HBase, biggest case), and for the biggest case, the inference reaches a
maximum of about~10x slower (MongoDB). This is expected as the query only
has to process a part of the database while the inference treats the whole
database. In the following sections, these results will be studied.

\begin{table}[htb!]
\resizebox{.49\textwidth}{!}{%
\begin{tabular}{@{}lrrrrr@{}}
    \toprule
    Size/Item & User & Movie & \parbox[b]{7em}{Watched/\\Favorite Movies} & Nodes & Relationships\\
    \midrule
    Larger & \numprint{800}k & \numprint{400}k & 20/user & \numprint{2000}k & \numprint{24800}k\\
    Large & \numprint{400}k & \numprint{200}k & 10/user & \numprint{1000}k & \numprint{6400}k\\
    Medium & \numprint{200}k & \numprint{100}k & 5/user & \numprint{500}k & \numprint{1700}k\\
    Small & \numprint{100}k & \numprint{50}k & 3/user & \numprint{250}k & \numprint{550}k\\
    \bottomrule
\end{tabular}%
}
\vspace*{1mm}
\caption{Database Sizes.\label{tab:database-sizes}}
\end{table}

\begin{figure}[htb!]
  \centering
  \includegraphics[width=\columnwidth]{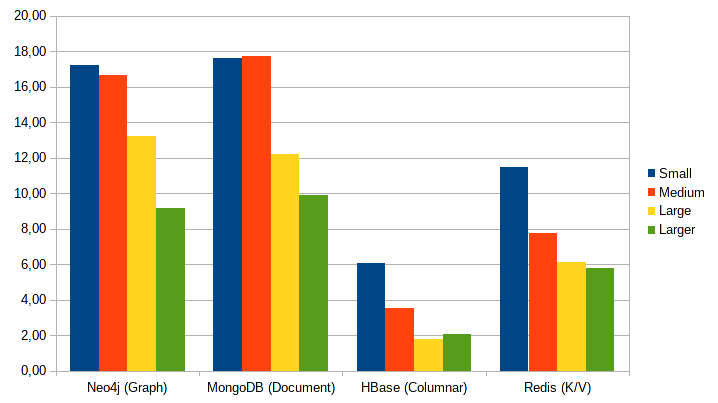}
  \caption{Inference to Query time ratio.}
  \label{fig:inference-ratio}
\end{figure}

\begin{table}[htb!]
\resizebox{.49\textwidth}{!}{%
\begin{tabular}{@{}llrrrr@{}}
\toprule
DB  &  & Small & Medium & Large & Larger\\
\midrule
Neo4j & Query (ms) & \numprint{686} & \numprint{1213} & \numprint{3165} & \numprint{12016}\\
 & Inference (ms) & \numprint{11821} & \numprint{20177} & \numprint{41814} & \numprint{109724}\\
 & Normalized & \numprint{17.23} & \numprint{16.63} & \numprint{13.21} & \numprint{9.13}\\
\midrule
MongoDB & Query (ms) & \numprint{295} & \numprint{380} & \numprint{840} & \numprint{2366}\\
 & Inference (ms) & \numprint{5187} & \numprint{6730} & \numprint{10226} & \numprint{23452}\\
 & Normalized & \numprint{17.58} & \numprint{17.71} & \numprint{12.17} & \numprint{9.91}\\
\midrule
HBase & Query (ms) & \numprint{931} & \numprint{1942} & \numprint{6419} & \numprint{24023}\\
 & Inference (ms) & \numprint{5615} & \numprint{6840} & \numprint{11526} & \numprint{49042}\\
 & Normalized & \numprint{6.03} & \numprint{3.52} & \numprint{1.80} & \numprint{2.04}\\
\midrule
Redis & Qyery (ms) & \numprint{1002} & \numprint{2833} & \numprint{10091} & \numprint{43888}\\
 & Inference (ms) & \numprint{11487} & \numprint{22013} & \numprint{61505} & \numprint{252794}\\
 & Normalized & \numprint{11.46} & \numprint{7.77} & \numprint{6.10} & \numprint{5.76}\\
\bottomrule
\end{tabular}%
}
\vspace*{1mm}
\caption{Times for inference and queries for all the database
  implementations.\label{tab:database-times}}
\end{table}

With the extracted \uschema{} model, we build a set of queries on the
databases to perform the second experiment:

\begin{enumerate}
\item {\bf All variations exist} The database must store, at least, a
  database object for each entity type variation (and relationship type
  variation in the case of a graph store) present in the extracted
  \uschema{} model.
\item {\bf Count correctness} No other variations are present in the
  database, i.e.,~the total number of objects in the database matches the
  sum of objects that belong to each structural variation of the entity
  types present in the extracted model (\texttt{count} attribute included
  in the \texttt{StructuralVariation} metaclass of the \uschema{}
  metamodel.) Also, this check would be performed for relationship type
  variations in the case of graph stores.
\end{enumerate}

\section{Representing Graph Databases as \uschema{}
  Models\label{sec:graph-database}}




\subsection{A Data Model for Graph Databases\label{graph-datamodel}}

In graph systems (e.g.,~Neo4j and OrientDB), a database is organized as a
graph whose nodes (a.k.a.~vertex) and edges (a.k.a.~arcs) are data items
that correspond to database entities and relationships between them,
respectively. Edges are directed from an \emph{origin node} to a
\emph{destination node}, and more than one edge can exist for the same pair
of nodes. Both nodes and edges can have \emph{labels} and
\emph{properties}. Labels denote the entity or relationship type to which
nodes or relationships belong, and properties are key-value pairs. This is
the so called \emph{labeled property graph data model}~\cite{angles2017},
that most NoSQL graph systems implement.

Graph databases are commonly schemaless, so there may exist nodes and
relationships with the same label but different set of properties.
Moreover, the same label can be used to name relationships that differ in
the type of the origin and/or destination nodes. Thus, graph databases can
have structural variations as explained in Section~\ref{sec:uschema-concepts}.

For this kind of graph store, we have abstracted the following notion of
logical \emph{graph data model}, which is represented in form of UML class diagram in
Figure~\ref{fig:GraphDataModel}:

\begin{figure*}[!htb]
  \centering
  \includegraphics[width=0.85\linewidth]{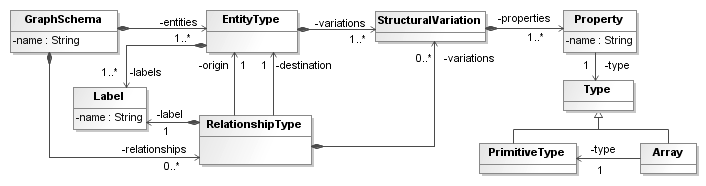}
  \caption{Graph Data Model.\label{fig:GraphDataModel}}
\end{figure*}

\begin{enumerate}
\item[i)] A graph schema has a name (that of the database) and is formed by
  a set of \emph{entity types} and a set of \emph{relationship types}.
\item[ii)] An entity type denotes the set of nodes with the same label (or
  set of labels).
\item[iii)] Entity types can be single-label or multi-label depending on
   whether they have one or more labels.
\item[iv)] A relationship type denotes the set of relationships with the same
  label (or set of labels). A relationship type has origin and destination
  entity types.
\item[v)] Entity and relationship types can have structural variations.
\item[vi)] A structural variation is characterized by a set of properties that
  is shared by elements with the same set of labels.
\item[vii)] A property is a pair that mimics the property of a node or
  relationship in the graph, having a key and the scalar data type that
  corresponds to the values of the property.
\end{enumerate}

Table~\ref{table:uschema-mappings} shows the correspondence between graph
database elements and the graph model elements expressed above. Note that a
graph schema is obtained by the MapReduce operation of the schema
extraction process described in the previous section.

\begin{figure}[!htb]
  \includegraphics[width=\linewidth]{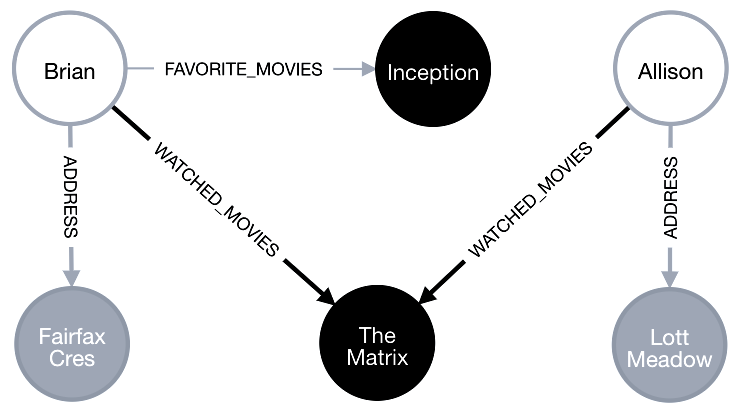}
  \caption{``User Profiles'' Graph Database
    Example.\label{fig:neo4j-ejemplo-grafo}}
\end{figure}

Figure~\ref{fig:neo4j-ejemplo-grafo} shows a graph database for the ``User
Profiles'' running example.
It has three entity types labeled \textit{Movie}, \textit{User}, and
\textit{Address}, and three relationship types labeled
\textit{FAVORITE\_MOVIES}, \textit{WATCHED\_MOVIES}, and \textit{ADDRESS}. In
the figure, nodes are represented as circles, and relationships as arrows.
Nodes having the same labels (i.e.~entity type) are filled with the same
color. In this example, gray for \textit{Address}, white for \textit{User},
and black for \textit{Movie}. Nodes only show a property for each entity
type: \textit{title} for \textit{Movie}, \textit{name} for \textit{User},
and \textit{street} for \textit{Address}. Relationships are tagged with
their relationship types, and no properties are shown. We suppose that
there are the variations indicated in Section~\ref{running-example}.

\subsection{Canonical Mapping between Graph Model and
  \uschema{}\label{sec:graph-mapping}}


Each element of the graph model defined above has a natural mapping to a
\uschema{} element, with the exception of relationship types, that map to
two \uschema{} elements: \texttt{RelationshipType} and \texttt{Reference}.
The former represents a type or classifier whose instances are
relationships between a pair of nodes, and can have variations based in
their set of attributes, while the latter denotes a particular link between
two nodes.
Note that \texttt{Aggregation} and \texttt{Key} \uschema{} elements do not
have a direct correspondence to elements of the graph model. Next, we
express the set of rules that defines the Graph to \uschema{} canonical
mapping.

\urule{R1}. A graph schema $G$ corresponds to an instance $uS$ of the
\texttt{uSchemaModel} metaclass of \texttt{\uschema{}} (i.e.,~a schema or
model) with the same name:
\[uS \leftrightarrow G\, \|\, \{uS.name = name(G)\}\]
\urule{R2}. Each different single-label entity type $e$ that exists in $G$
maps to a root \texttt{EntityType} $et$ in the $uS$ schema, whose name is
that of the label associated to $e$:
\[et \leftrightarrow e \| \{et.name = name(e),\, et.root \leftarrow
  true\}\] \texttt{EntityType} instances are included in the $uS.entities$
collection.

\urule{R3}. Each different multiple-label entity type $e$ that exists in
$G$ maps to a root \texttt{EntityType} $et$ in the $uS$ schema whose name
is formed by concatenating the names of the set of $n>1$ labels
$L=\{l_1,\dots,l_n\}$, and $et$ inherits from each entity type
${e_1,\dots,e_n}$ that corresponds to labels in $L$:
\[\begin{aligned}
    et \leftrightarrow e & \,\|\, \{et.name = concat(L), \\
    &et.root \leftarrow true,\, \\
               & et.parents = set \{map(e_1),\dots,map(e_n)\}\}
\end{aligned}\]


\urule{R4}. Each relationship type $r$ that exists in $G$ maps to
a \texttt{RelationshipType} $rt$ and a \texttt{Reference} $rf$ in the $uS$
schema, which are named the same as the label associated to $r$.
\[r \leftrightarrow rt\, \|\, \{rt.name = name(r)\}, \]
\[r \leftrightarrow rf\, \|\, \{rf.name = name(r)\}\]
\texttt{RelationshipType} instances are included in the $uS.relationships$
collection, and Rule~R7 specifies how references are connected to other
elements of the \uschema{} schema.

\urule{R5}. Each ``variation schema'' $v$ of an entity or
relationship type in $G$ maps to a \texttt{StructuralVariation} $sv$ in the
$uS$ schema, which is identified by means of a unique identifier $index$
(an integer ranging from~1 to $|EV|$ or $|RV|$). Structural variations are
included in the collection $variations$ that both entity types and
relationship types have in a \uschema{} schema.

\urule{R6}. Let $P^v$ be the set of properties of a ``variation
schema'' $v$ which maps to a \texttt{StructuralVariation} $sv$. Each
property $p^v_i \in P^v$ will map to an \texttt{Attribute} $at^{sv}_{i}$
with the same name, which is included in the collection $sv.features$. The
type of the property will map to one of the types defined in the
\texttt{Type} hierarchy defined in \uschema{}, and a mapping has to be
specified for each graph store. The property mapping can be expressed as:
\[\begin{aligned}
  p^{v}_{i} \leftrightarrow at^{sv}_{i}\, \|\, \{at^{sv}_{i}.name =
  name(p^{v}_{i}),\, \\
  at^{sv}_{i}.type \leftrightarrow type(p^{v}_{i}) \}
\end{aligned}\]

\urule{R7}. Each reference in a \uschema{} schema $uS$ has to be connected
to other elements of $uS$. Let $rf$ be a \texttt{Reference} which maps to a
relationship type $r$ according to Rule~\textbf{R4},

\begin{itemize}
\item[i)] $rf$ must be linked to the \texttt{EntityType} that maps to the
  entity type that denotes the destination nodes for the relationship
  $r$:~$rf.refs\_to \leftarrow map(destination(r))$.

\item[ii)] Let $oe$ the {\tt EntityType} of $uS$ that maps to the origin
  entity type of a relationship type $r$ in $G$
  ($oe \leftrightarrow map(origin(r))$), $rf$ will be present in the set of
  features of the variations of $oe$ whose nodes are origin of edges that
  are instances of $r$.


\item[iii)] $rf$ must be linked to the structural variation which
  features: $rf.isFeaturedBy \leftarrow sv$, where $sv$ is the
  \texttt{StructuralVariation} that belongs to the relationship type that
  returns $map(r, RelationshipType)$.

\item[iv)] The \texttt{lowerbound} cardinality of $rf$ would be $1$
  ($rf.lowerBound \leftarrow 1$) and the \texttt{upperbound} cardinality
  could be $1$ ($ref.upperBound \leftarrow 1$) or $\infty$
  ($ref.upperBound \leftarrow \infty $) depending on whether the instances
  of $r$ in the database (i.e. arcs of type $r$)
  have one or more destination nodes for a given origin node.
\end{itemize}

\subsection{Reverse Mapping Completeness\label{sec:reverse-mapping-graph}}

The graph model does not include the \texttt{Key} and \texttt{Aggregate}
elements. Next, we provide a possible mapping for these two concepts.

\begin{figure}[!h]
  \centering
  \includegraphics[width=0.8\linewidth]{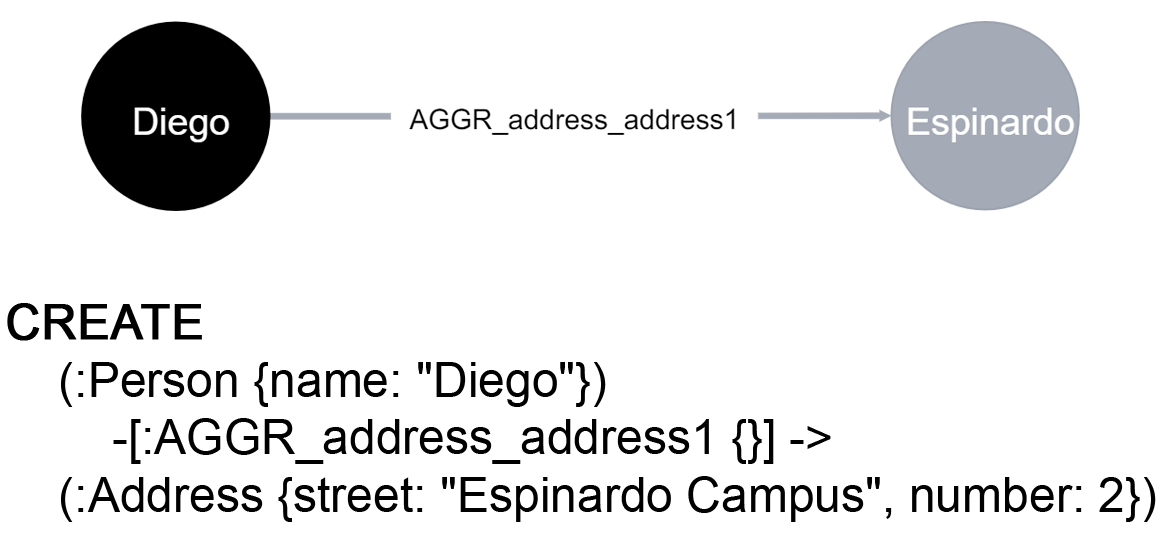}
  \caption{Person Aggregates Address in
    Neo4j.\label{fig:neo4jpersonaggregate}}
\end{figure}

\begin{itemize}
\item \textbf{Key}. Remember that the \texttt{Key} concept in \uschema{}
  refers to those attributes that act as an object key or the set of
  attributes that form part of a reference to another object. As references
  between objects (nodes) in graphs are explicit in arcs, there is no need
  to include key information into the graph schema. However, that
  information could be included in the nodes, for example, using a special
  $\_keys$ property holding the set of properties that act as key.

\item \textbf{Aggregate}. An \texttt{Aggregate} $ag$ that belongs to a
  particular \texttt{StructuralVariation} $sv$, where $ag.aggregates$ is
  the aggregated variation $av$, could be mapped to a relationship type
  whose name is $ag.name$ adding the prefix ``\texttt{AGGR\_}'', its origin
  entity type being $sv.container$, and its destination entity type being
  $av.container$. Origin and destination entity types should be created if
  they do not exist in the graph schema. Also, properties of $av$ should be
  mapped using rules~R2 to~R7, as well as this rule (if an aggregate is
  part of the properties of $av$). Figure~\ref{fig:persondata} shows an
  example JSON document of an \uschema{} entity type {\tt Person} that
  aggregates an object of the entity type {\tt Address}.
  Figure~\ref{fig:neo4jpersonaggregate} illustrates the reverse mapping
  where a relationship type named {\tt AGGR\_address\_address1} connects a
  {\tt Person} and {\tt Address} nodes (we suppose that the aggregated
  variation identifier is~1.)


\end{itemize}

\begin{figure}[!ht]
\centering
\begin{minipage}[t]{.45\textwidth}
\begin{lstlisting}[language=json]
Person:
{
 name: "Diego",
 address: { street: "Espinardo Campus", number: 2}
}
\end{lstlisting}
\end{minipage}
\caption{Example Person Data with Address Aggregate.\label{fig:persondata}}
\end{figure}

\subsection{Implementation and Validation of the Forward
  Mapping for Neo4j\label{sec:graph-mapping-implementation}}

A slightly revised strategy to that described in
Section~\ref{sec:commonstrategy} has been applied to implement the forward
canonical mapping for Neo4j. We chose this store because it is the most
popular graph database.\footnote{DB-Engines Ranking
  \url{https://db-engines.com/en/ranking}, (January,~2021).} It is
schemaless and fits into the \emph{labeled property graph data model}.

The strategy had to be revised because graph databases usually do not offer
facilities to efficiently process the whole graph, and sometimes they even
fail because of lack of resources. So we devised a preliminary stage that
serialized the graph obtaining all the nodes along with their outgoing
relationships. Of each arch, the data included the source node with its
properties, the properties of the arc, and the ID of the destination node.
We modified the \emph{map} operation of the generic strategy to construct
all the raw schemas for nodes and edges with this serialization format. The
serialization was organized in batches by using Spark Neo4j
connector~\cite{spark-web}. This way, an efficient schema extraction
process was achieved. The reduce operation did not need any modification
from that described in the generic strategy, generating variation schemas
for both entity and relationship types from nodes and arcs, respectively.

\begin{figure*}[!htb]
   \begin{subfigure}[c]{0.55\textwidth}
      \includegraphics[width=\textwidth]{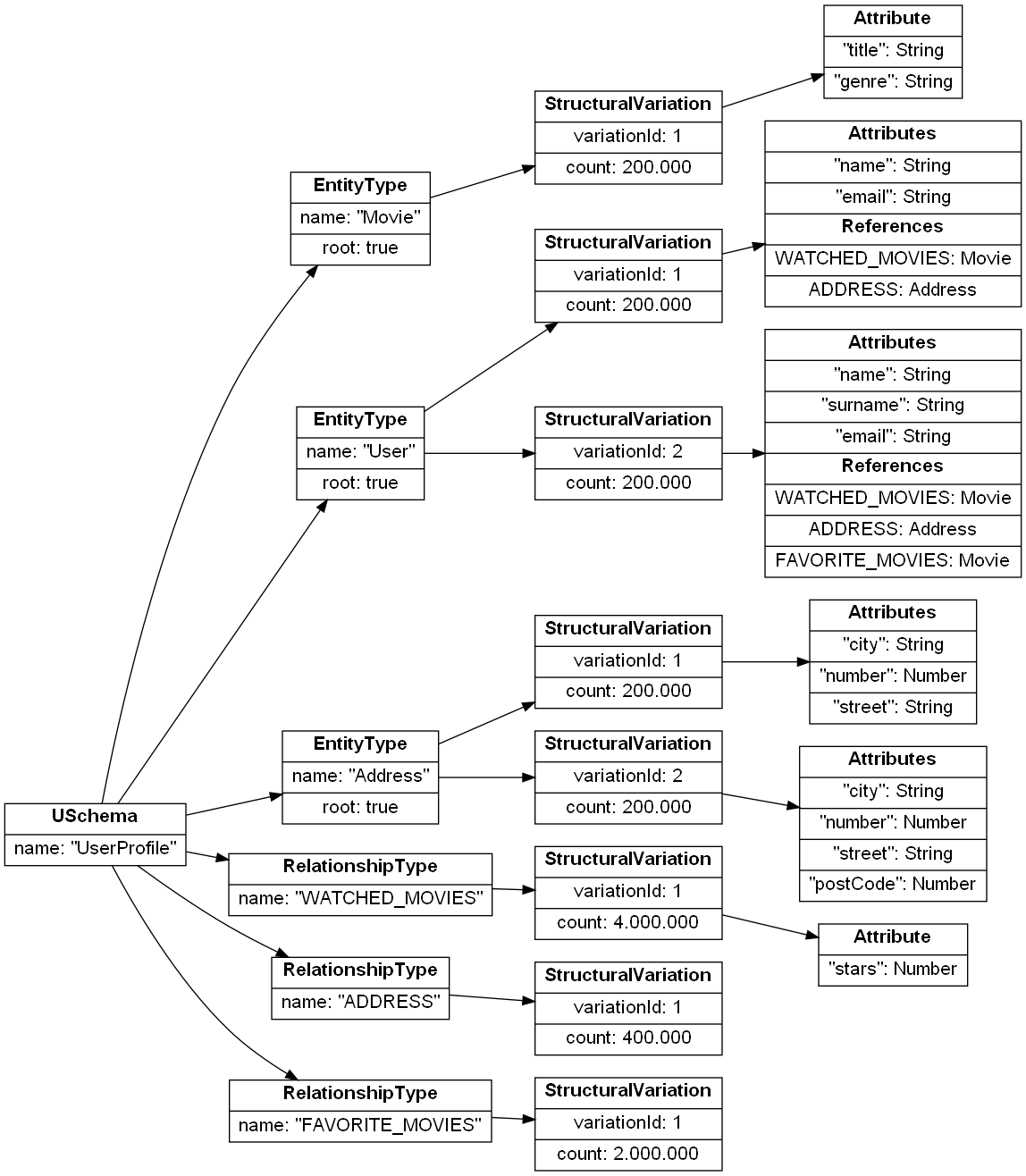}
      \caption{Complete Schema.\label{fig:schema_graph_a}}
   \end{subfigure}%
   \begin{subfigure}[c]{0.45\textwidth}
      \includegraphics[width=\textwidth]{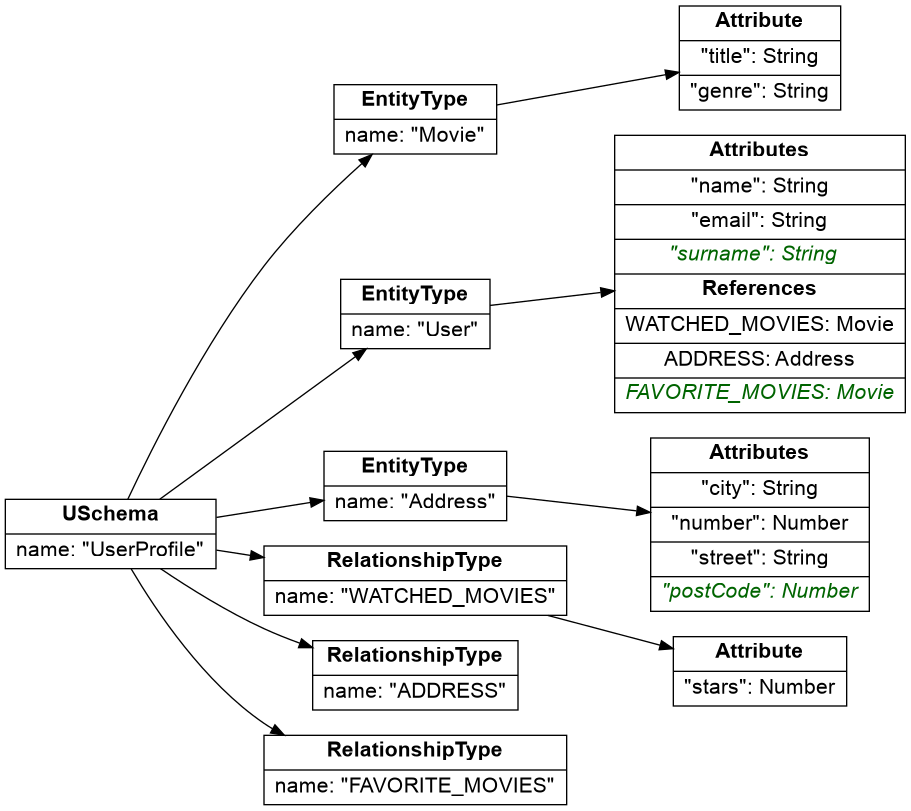}
      \caption{Union Schema.\label{fig:union_schema_graph}}
    \end{subfigure}
    \caption{``User Profiles'' Schema for Graph
      Stores.\label{fig:schema_graph}}
\end{figure*}

The process finalizes with creating the \uschema{} model by applying the
mapping rules to the previous output (i.e.~the \emph{logical graph model}).
The resulting schema for the ``User Profiles'' running example is shown in
Figure~\ref{fig:schema_graph_a}. We also show the \emph{union schema} in
Figure~\ref{fig:union_schema_graph}.




The two experiments introduced in Section~\ref{common-validation} were
successfully carried out on the Neo4j database created for the running
example and a \emph{Movies} dataset available at the Neo4j
website.\footnote{No longer available at the original site, a copy can be
  obtained from
  \url{https://github.com/catedrasaes-umu/NoSQLDataEngineering/blob/master/data/Neo4j/Movies/}.}

Regarding scalability and efficiency of the model creation process,
Table~\ref{tab:database-times} show that the relative times with the
reference query decrease as the size of the database increases. Neo4j,
jointly with MongoDB show the worst ratio cases. This is because the query
(average of watched movies by user) is, by chance, easily optimized by the
database. In any case, as the database grows, the factor is never
beyond~10x.

\section{Representing Document Databases as \uschema{}
  Models\label{sec:document-database}}

\subsection{A Data Model for Document
  Databases\label{sec:document-datamodel}}

Document databases (e.g., MongoDB and Couchbase) are organized in
collections of data recorded for a particular database entity
(e.g.,~\emph{Movie}, \emph{User}, and \emph{Address} in the running
example). Data are stored in the form of semi-structured objects or
documents~\cite{abiteboul-data2000,buneman1997} that consist of a tuple of
key-value pairs (a.k.a.~fields). Keys denote properties or attributes of
the entity, and the values can be atomic data (e.g.~Number, String, or
Boolean), nested or embedded documents, or an array of values. Also, a
string or integer value can act as a reference to another document, similar
to foreign keys in relational systems, although usually no support for
consistency is provided.



Semi-structured data is characterized by having its schema implicit in
itself~\cite{buneman1997}. Thus, document databases are commonly
schemaless, and a collection can store different \emph{variations} of the
entity documents. Usually, document databases maintain data in some
JSON-like format.

\begin{figure*}[!htb]
  \centering
  \includegraphics[width=0.85\linewidth]{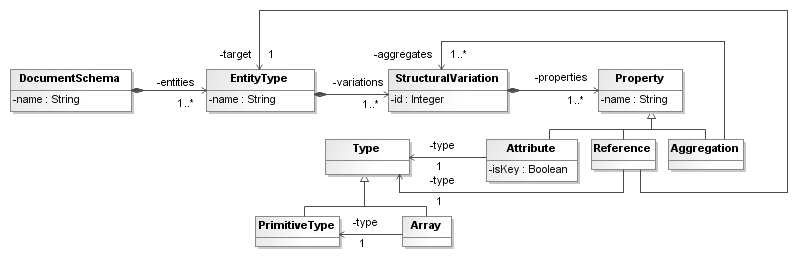}
  \caption{Document Data Model.\label{fig:DocumentDataModel}}
\end{figure*}

For document databases, we have abstracted the following notion of
\emph{document data model}, which is represented in form of a UML class
diagram in Figure~\ref{fig:DocumentDataModel}:

\begin{enumerate}
\item[i)] A document schema has a name (that of the database) and is formed
  by a set of \emph{entity types}.
\item[ii)] An entity type denotes a collection of documents stored in the
  database.
\item[iii)] Entity types have one or more structural variations.
\item[iv)] A structural variation is characterized by a set of properties
  that are shared by documents of the same collection.
\item[v)] Properties have a name and a type, and can be attributes,
  aggregates, or references.
\item[vi)] Attributes denote object's fields whose value is of scalar or
  array type. An attribute is specified by the name of the field and the
  type of its value. We suppose that there exists an attribute that acts as
  the key of the Entity type (e.g.,~``\texttt{\_id}'' in MongoDB).
\item[vii)] Aggregates denote object's fields whose value is an embedded
  object. An aggregate is specified by the name of the field and the
  variation schema of the embedded object.
\item[viii)] References denote object's fields whose values are references.
  A reference is specified by the name of the field and the type of its
  value.
\end{enumerate}

Table~\ref{table:uschema-mappings} shows the correspondence between
document database elements and the document model elements expressed above.
Note that a document schema is obtained by the MapReduce operation of the
schema extraction process described in Section~\ref{sec:commonstrategy}.

\begin{figure*}[!htb]
  \centering
  \includegraphics[width=.8\textwidth]{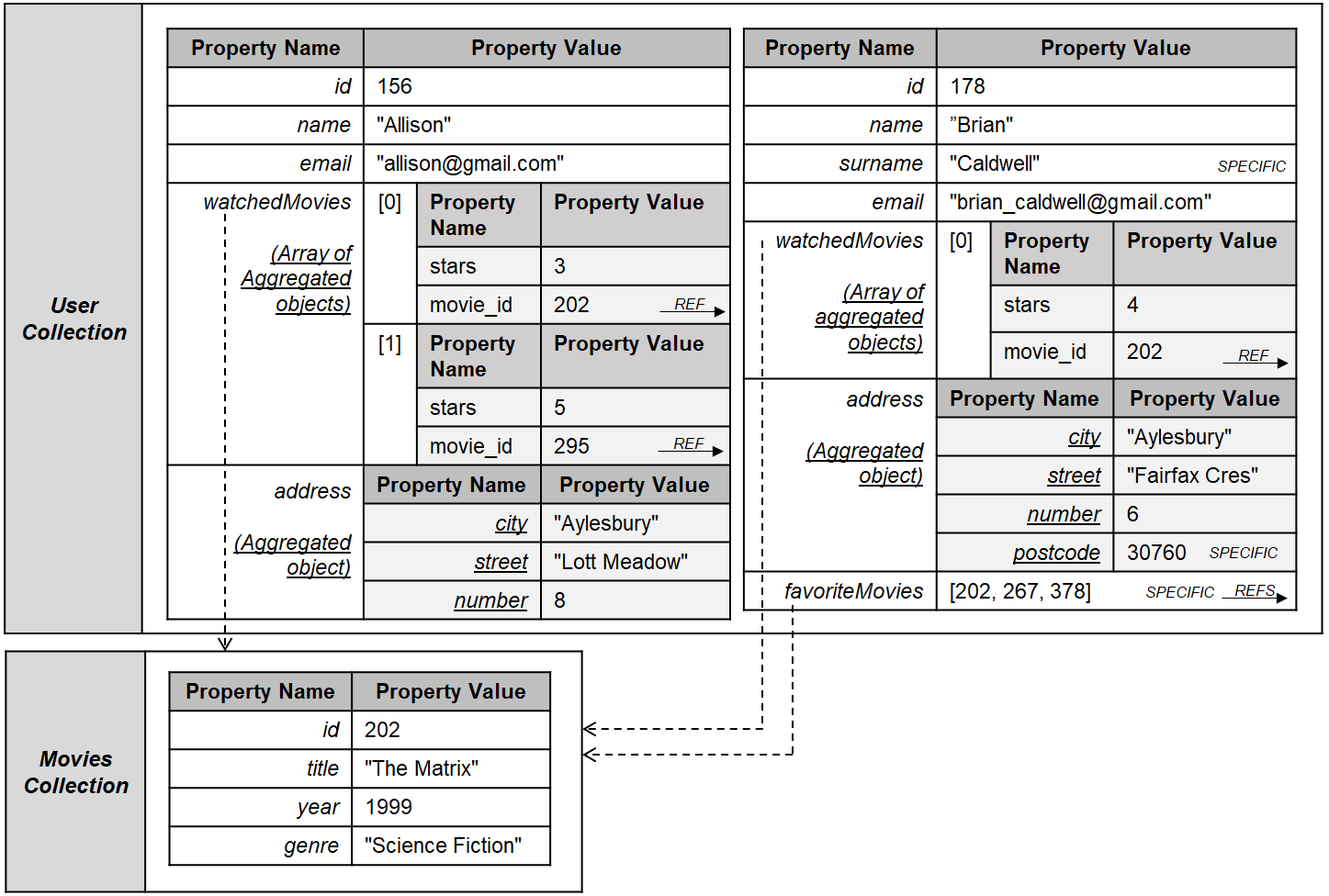}
  \caption{``User Profiles'' Document Database
    Example.\label{fig:document-database-example}}
\end{figure*}

Figure~\ref{fig:document-database-example} shows how the ``User Profiles''
running example would be stored in a document database. Instead of using
JSON notation, we depicted the database objects in a representation
that remarks their nested structure and the references between objects.
There are two collections: \emph{User} and \emph{Movie} objects, and the
relationships are as follows. \textit{User} objects aggregate
\emph{watchedMovies} objects with two properties: the \textit{stars}
attribute and the \textit{movie\_id} reference that records the \textit{id}
value of a movie object (arrow from \textit{movie\_id} to \emph{Movie}
objects); \emph{watchedMovies} objects are recorded in an array.
To record favorite movies, \textit{User} has the \textit{favoriteMovies}
array of references to \textit{Movie} objects. The user addresses are
stored as an \textit{address} aggregate object of users. While graph
databases rely on references (i.e.~relationships in graph store
terminology) to connect data items, and aggregation is normally not
available to compose data, the opposite is true in document database
systems.

\subsection{Canonical Mapping between Document Model and
  \uschema{}\label{sec:document-mapping}}

Each element of the document data model defined above has a natural mapping
to a \uschema{} element. Next, we present the rules for the canonical
mapping.




\urule{R1}. A document schema $D$ corresponds to an instance $uS$ of the
\texttt{uSchemaModel} metaclass of \texttt{\uschema{}} (i.e.,~a schema or
model) with the same name:
\[uS \leftrightarrow D\, \|\, \{uS.name = name(D)\}\]

\urule{R2}. Each entity type $e$ that exists in $D$ maps to a root
\texttt{EntityType} $et$ with the same name:
\[et \leftrightarrow e \| \{et.name = name(e) ,\, et.root \leftarrow true\}\]
$uS.entities$ holds the set of instances of \texttt{EntityType}.

\urule{R3}. Each variation schema $v$ of $e$ corresponds to a
\texttt{StructuralVariation} $sv$ of $et$ in the $uS$ schema, which is
identified by means of a unique identifier $index$ (an integer ranging
from~1 to $|EV|$). Each property $p^v_i$ of $v$ will be mapped according to
rules R4--R6.
\[\begin{aligned}
   sv \leftrightarrow v & \,\|\, \{sv.variationId = idgen(), \\
        &sv.features \leftrightarrow properties(v) \}
\end{aligned}
\]
\texttt{StructuralVariation} instances are included in the collection
$et.variations$.

\urule{R4}. If $p^v_i$ is an attribute,
\begin{itemize}
\item[i)] it will map to an \texttt{Attribute} $at^{sv}_i$ with the same
  name, which is included in the collection $sv.features$. The mapping is
  the same as that defined in Rule~\textbf{R6} of the mapping between the
  graph model and \uschema{}.
\item[ii)] Additionally, if the attribute is the key of the entity type, a
  {\tt Key} instance also exists in $sv.features$ and is connected to the
  corresponding attribute $at^{sv}_i$.
\end{itemize}

\urule{R5}. If $p^v_i$ is an aggregate that has a set of $n$ properties
$G^v=\{g^v_i\}, i=1..n$, it will map to three elements in the \uschema{}
model:

\begin{itemize}
\item[i)] A non-root \texttt{EntityType} $nr$ with the same name but
  capitalized and stemmed (function $name^*()$), which is included in the
  collection $uS.entities$:
  \[ nr \leftrightarrow p^v_i \, \|\, \{nr.name = name^*(p^v_i), \, nr.root
  \leftarrow false\}\]

\item[ii)] A \texttt{StructuralVariation} instance $av$ included in
  $nr.variations$, and each property $g^v_i$ is mapped recursively
  according to rules~R4 to~R6:
  \[av \leftrightarrow p^v_i \, \|\, \{ av.features \leftrightarrow G^v  \}\]

\item[iii)] An \texttt{Aggregate} $ag$ with the same name as the property,
  which is included in $sv.features$. This aggregate $ag$ is connected to
  the structural variation $av$ that it aggregates. The mapping would be:
  \[\begin{aligned}
  ag \leftrightarrow  p^v_i  \, \|\, \{ ag.name = name(p^v_i), \, \\
    av \in ag.aggregates\}
  \end{aligned}\] The cardinality of $ag$ would be established as indicated
in Rule~R7-ii of the mapping between graph models and \uschema{} models.
\end{itemize}

\urule{R6}. If $p^v_i$ is a reference, it corresponds to two elements of
the \uschema{} model:

\begin{itemize}
\item[i)] A \texttt{Reference} $rf$ with the same name, which is included in
  $sv.features$. The mapping is the same defined in Rule~R7 of the
  mapping between graph models and \uschema{} models.

\item[ii)] An \texttt{Attribute} $at$ according to the mapping expressed in
  Rule~R4-i, and $at$ and $rf$ appear connected in the $uS$ schema:
  $at$ exists in $rf.attributes$ and $rf$ is part of $at.references$.
\end{itemize}

\subsection{Reverse Mapping Completeness \label{reverse-document}}

The only element of \uschema{} that is not directly supported by the
document model is the {\tt RelationshipType}. {\tt RelationshipType}s
have structural variations, and some {\tt Reference}s can specify (via its
{\tt isFeaturedBy} property) to which {\tt StructuralVariation} of a {\tt
  RelationshipType} they belong.

Given a {\tt RelationshipType} $rt$ in a \uschema{} model, the reverse
mapping for documents would map to an entity type $e$ whose name is
$rt.name+"\_REF"$. Each variation of $rt$ will correspond to a variation in
$e$, applying mapping rule R3 (i.e.,~each {\tt Attribute} in $rt$ maps to
an attribute of the corresponding variation of $e$). A reference property
$p$ will exist in all the variations of $e$ that will map with rule R6.
Then, each {\tt Reference} $rf$ that belongs to a {\tt StructuralVariation}
$v$ of the entity type $et$ to which $origin(rt)$ maps, where $ro$ is the
relationship type such that $ro=map(rt)$, and whose {\tt isFeaturedBy} is a
variation $vt$ in $rt.variations$, will map to a reference property $r$
named $name(e)+"\_ref"$ by applying rule~R6.

\[\begin{aligned}%
    rf \leftrightarrow r \| \{&rf \in v.features,\\
    &et \leftrightarrow origin(rt),\\
    &v \in map(et).variations, \\
    & vt = rf.isFeaturedBy, \\
    & vt \in rt.variations , \\
    & r.name \leftarrow name(e)+"\_ref"\}%
\end{aligned}%
\]

\begin{figure*}[!htb]
  \begin{subfigure}[c]{.38\textwidth}
    \includegraphics[width=\linewidth]{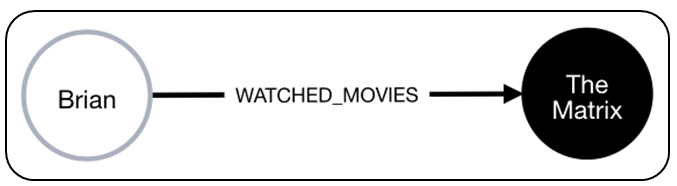}
    Documents:\\
    \vspace*{-1em}
\begin{lstlisting}[language=JSON]
// User Collection
{
  "name": "Brian",
  ...
  "watchedMovies": [
        978
  ]
}

// WatchedMovie_REF Collection
{
  "_id": 978,
  "stars": 3,
  "movie_id": 202
}

// Movie Collection
{
  "_id": 202,
  "title": "The Matrix"
  ...
}
\end{lstlisting}
  \end{subfigure}
  \begin{subfigure}[c]{.62\textwidth}
    \includegraphics[width=\linewidth]{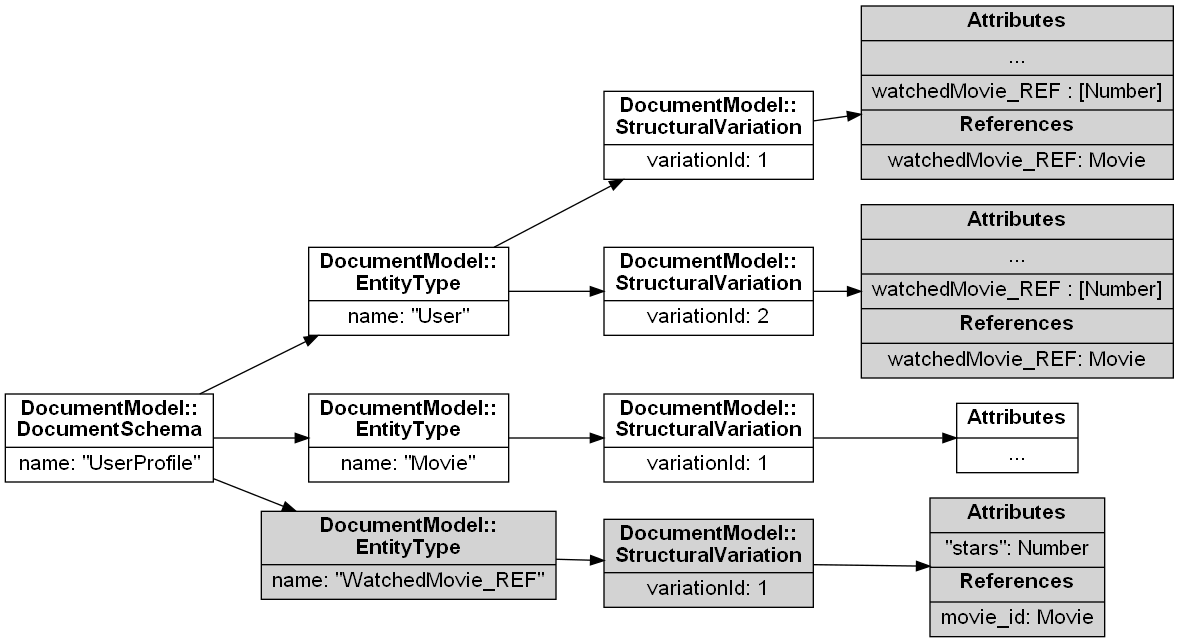}
  \end{subfigure}
  \caption{Example of Application of the Reverse Mapping from a
  \texttt{RelationshipType} of \uschema{} to a Document
  Schema.\label{fig:ReverseMappingRelationshipType}}
\end{figure*}

Figure~\ref{fig:ReverseMappingRelationshipType} illustrates the application
of the reverse mapping explained above for a \uschema{} model containing a
\texttt{RelationshipType} for the \emph{watchedMovies} relationship type of
the running example. It can be appreciated how the document schema would
contain an entity type named \emph{WatchedMovie\_REF}, which has a
structural variation for the single \texttt{StructuralVariation} of the
\texttt{RelationshipType} that exists in the \uschema{} schema. That
variation is connected to the attributes named \textit{stars} and
\textit{movie\_id}. Also, there exists a reference to the entity type
\emph{Movie}, and a reference and attribute named \emph{watchedMovie\_REF}
are present in the structural variations of the origin entity type
(\emph{User} in our example). The reference will connect the \emph{User}
objects with the \emph{WatchedMovie\_REF} objects.

Some document systems provide the \emph{dbref} construct to record
references between documents, which can include fields. In these systems,
the document data model shown in Figure~\ref{fig:DocumentDataModel} could
be extended to consider that references can have attributes. Then, the
document model would include all the \uschema{} elements, as it would also
support relationship types.

\subsection{Implementation and Validation of the Forward Mapping for
  MongoDB\label{sec:document-mapping-implementation}}

Once the output of the MapReduce described in
Section~\ref{common-implementation} is produced, i.e.,~the set of variation
schemas, the generation of the \uschema{} model is achieved by following
the mapping rules described above. The only remarkable aspect is that while
the root entity types are discovered by the MapReduce process, aggregated
entity types reside ``unfolded'' inside the variation schemas. It is needed
to recursively process all the aggregated objects to build the non-root
\texttt{EntityType}s and match the properties to identify the
\texttt{StructuralVariation}s.

\begin{figure*}[!htb]
   \begin{subfigure}[c]{.62\textwidth}
      \includegraphics[width=\textwidth]{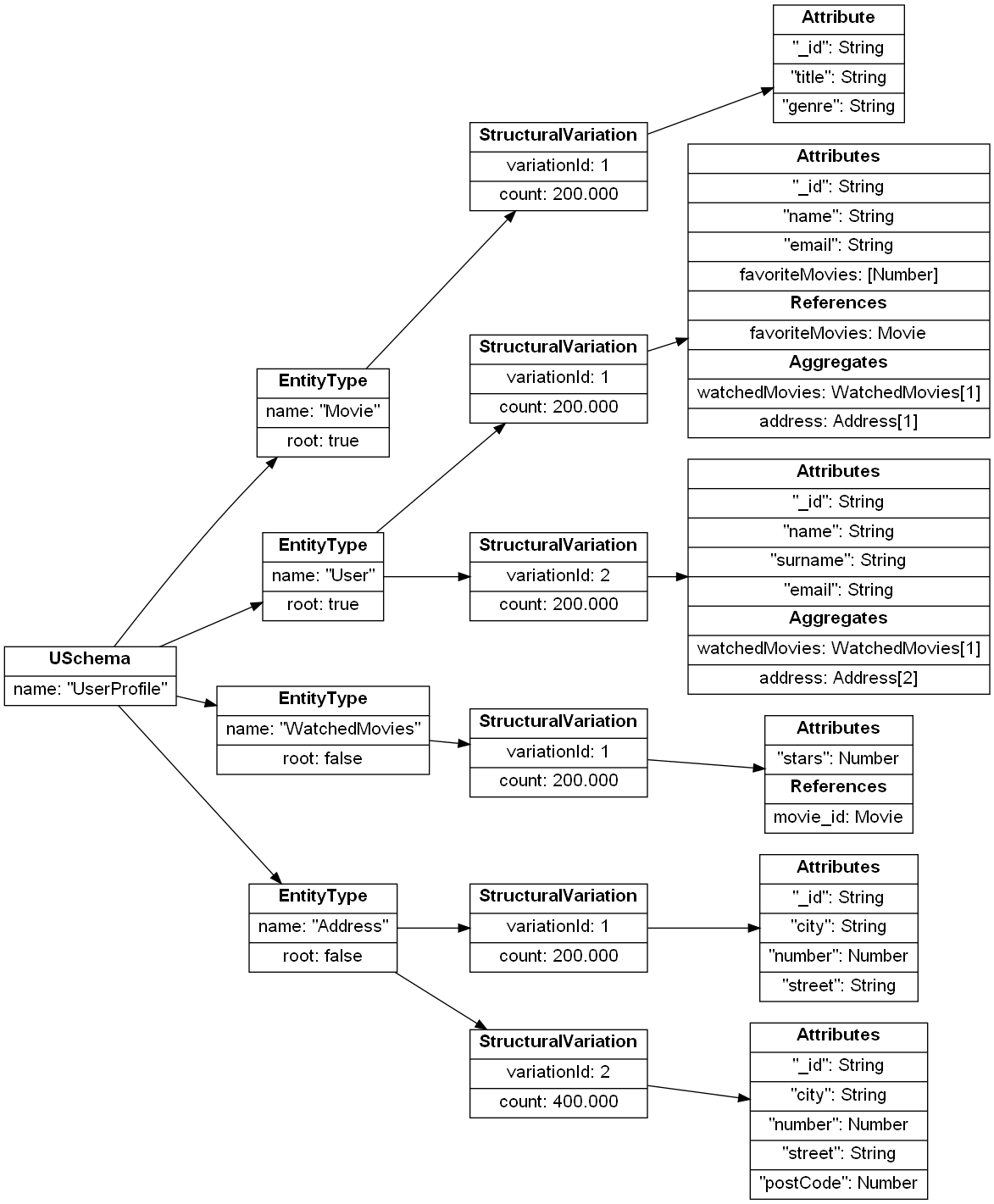}
      \caption{Complete Schema.\label{fig:schema_doc_kv_columnar1}}
   \end{subfigure}%
   \begin{subfigure}[c]{.4\textwidth}
  \includegraphics[width=\textwidth]{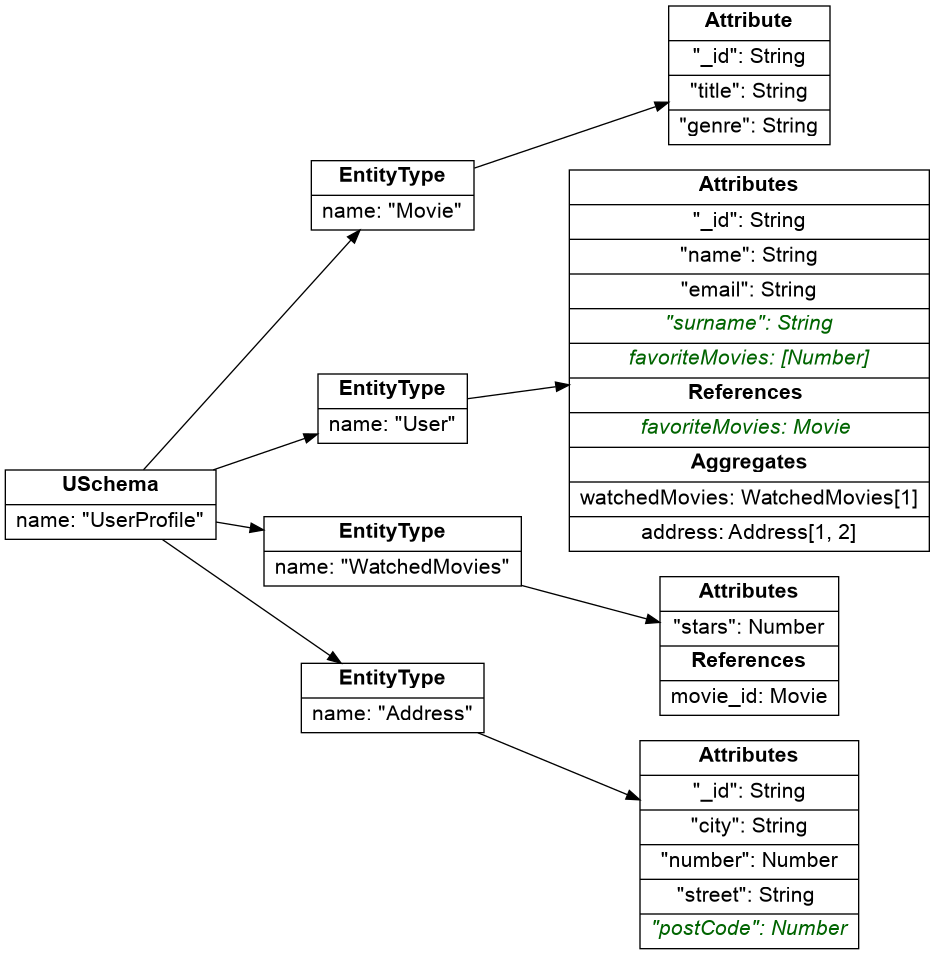}
      \caption{Union Schema.\label{fig:union_schema_doc_kv_columnar}}
   \end{subfigure}
   \caption{``User Profiles'' Schema and Union Schema for Document
     Stores.\label{fig:schema_doc_kv_columnar}}
\end{figure*}

The schema that would be inferred for the running example is shown in
Figure~\ref{fig:schema_doc_kv_columnar1} and the union schema in
Figure~\ref{fig:union_schema_doc_kv_columnar}.

The common validation strategy of Section~\ref{common-validation} was
successfully applied in MongoDB, with a database created for the running
example, and with the \emph{EveryPolitician} dataset.\footnote{Available at
  \url{http://docs.everypolitician.org/}.}

As with Neo4j, MongoDB shows worse ratio cases than with other two database
implementations, as shown in Table~\ref{tab:database-times}. Again, this
may be caused by the chance that the query benefits by some optimizations
built in the database. The ratio also goes down as the size of the database
doubles, with the exception of the Small and Medium times, that are similar
(\numprint{17.58}x and~\numprint{17.71}x). The ratio then goes down from
around~18x to~10x for the biggest case.

\section{Representing Key-Value Databases as \uschema{}
  Models\label{sec:keyvalue-inference}}

\subsection{A Data Model for Key-Value databases}

Key-Value (K/V) stores conform to the simplest physical data model of NoSQL
systems. A K/V store is an associative array, dictionary, map, or
\emph{keyspace}, that holds a set of key-value pairs, usually
lexicographically ordered by key. As such, they are used to record data
with a simple structure, and references and aggregations are not primitive
constructs to build up data. They usually store a single entity type
(e.g.~user profile, user login, or a shopping cart), although data of
several entity types could co-exist in the same keyspace.


Like document and columnar systems, K/V stores can record semi-structured
objects. Several techniques can be used to encode a tree-like structure
into key-value pairs, which use normally \emph{namespaces} to build
hierarchical key values. We chose one of the most commonly used encoding
patterns\footnote{\url{https://redislabs.com/redis-best-practices/data-storage-patterns/object-hash-storage/}.}
to define the canonical mapping between K/V databases and \uschema{}, to
which we will call the \emph{flattened key pattern of compound objects} or
simply \emph{flattened object-key pattern}.

\begin{figure*}[!htb]
  \centering
  \includegraphics[width=0.8\textwidth]{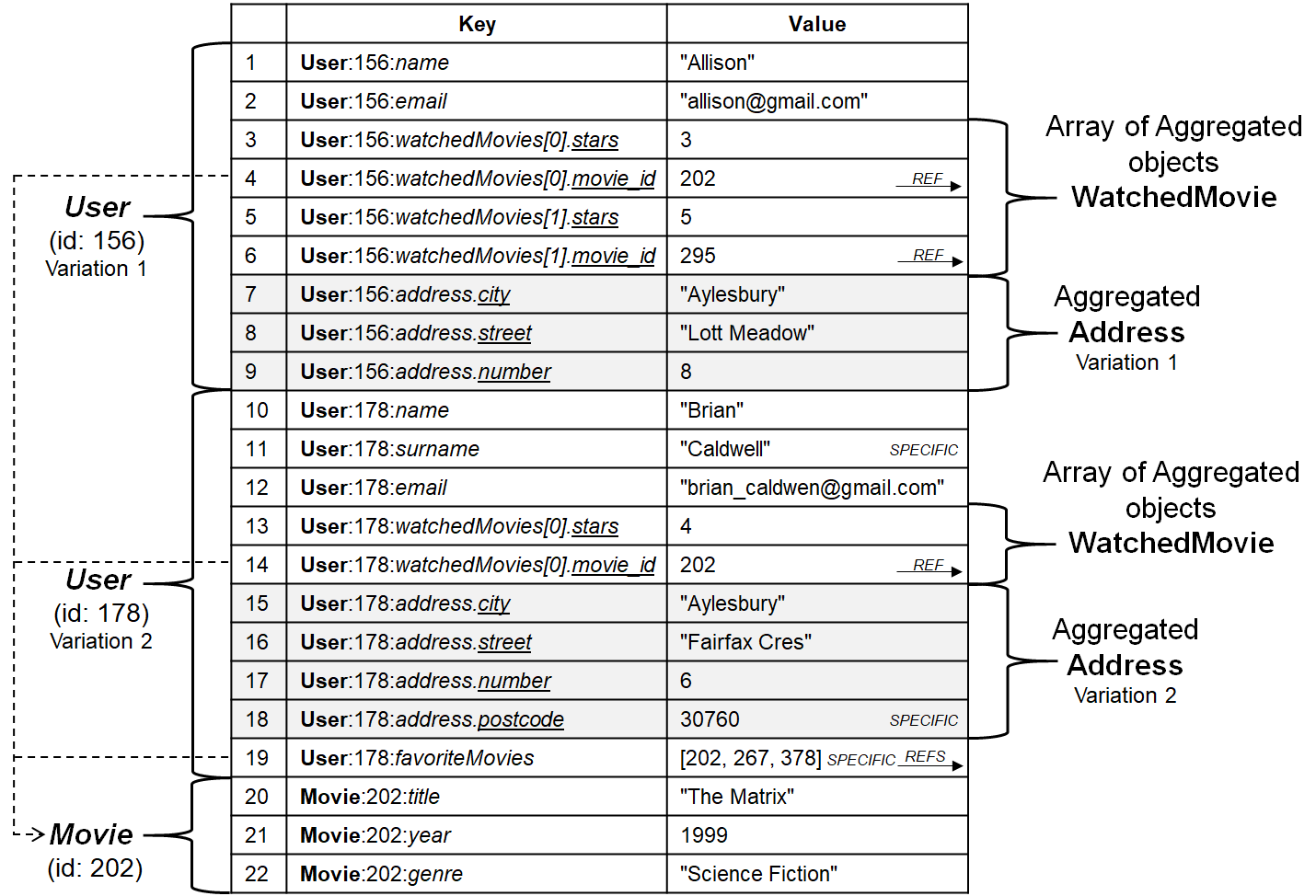}
  \caption{Key-Value Database Example for running
    example.\label{fig:kv-example}}
\end{figure*}

When using this pattern, the key of every pair not only acts as the
identifier of the object, but also encodes the name of a property of the
entity type, in a similar format to XPath or
JSONPath~\cite{gossner20:jsonp}. Keys are built with a separator to
differentiate between the object identifier and the property name (e.g.,~a
colon: ``\texttt{<id>:<property>}''). It can also be used to differentiate
the entity type if different namespaces are not used
(e.g.,~``\texttt{<entity-type>:<id>:<property>}''). When a property
aggregates an object, it is possible to use another separator to express
properties of the aggregated object (e.g.,~a dot:
``\texttt{<id>:<property>.<aggregated-property>}''), or an index to
represent objects of an array
(e.g.,~``\texttt{<id>:<property>[<index>]}''), that in turn can have
properties (e.g.,
``\texttt{<id>:<property>[<index>].<aggregated-property>}'').
Figure~\ref{fig:kv-example} shows an K/V database example that illustrates
the usage of this encoding for 
the ``User Profile'' running example. Using this pattern, a database object
consists of several entries in the database, all of them sharing the same
object identifier. Note that the order of the separated elements of the key
may vary depending on the specific queries needed by the application, as
the keys are lexicographically ordered.

K/V systems are schemaless, and several structural variations of an entity
type can therefore exist in the database. In Figure~\ref{fig:kv-example},
the variations of the running example can be observed.

Taking into account the use of the \emph{flattened object-key pattern}, the
document data model presented in Section~\ref{sec:document-datamodel}, and
shown in Figure~\ref{fig:DocumentDataModel}, can also be used for K/V
systems by modifying the \emph{Key} notion. In this case, every database
object also has a key, but it is not associated to any attribute. A
namespace would correspond to an entity type or either, if only one
namespace is used, each different entity type will have a different
``\texttt{<entity-type>}'' key prefix. This data model, as the Document
model, has all the elements of \uschema{} except the
\texttt{RelationshipType} element, as shown in
Table~\ref{table:uschema-mappings}.
We will use the term \emph{aggregate-oriented data model} to group the
Document, Key-Value, and Columnar data models, as suggested
in~\cite{fowler-nosql2012}, because they include the same concepts in their
respective data models.

A set of data types are available for keys and values, which vary on each
system. Keys are normally stored as byte-arrays or strings, which can
follow formats as those indicated above. Regarding the data types of
values, they usually provide basic scalar types as well as common
collection types.

\subsection{Canonical Mapping Between Key-Value Model and
  \uschema{}\label{sec:keyvalue-mapping}}

The mapping between \uschema{} and Document model would be applicable for
K/V, the only exception being that Rule R4-ii should be removed, and a new
rule has to be added because the notion of key is different in this data
model.




\urule{R7}. Each {\tt StructuralVariation} $sv$ in the \uschema{} model
contains a \texttt{Key} instance $k$ in $sv.features$ whose value of
$k.name$ is ``\textit{\_id}'', and it is not connected to any
\texttt{Attribute}.


\subsection{Reverse Mapping Completeness}

The same reverse completeness mapping rules exposed in
Section~\ref{reverse-document} for the document model are applicable in
this case.

\subsection{Implementation and Validation of the Forward Mapping for
  Redis\label{sec:redis-mapping-implementation}}

Redis has been used for the implementation and validation of the general
strategy applied for key-value stores. Redis is the most popular key-value
database\footnote{As shown in \url{https://db-engines.com/en/ranking}.
  Redis appears in the~7$^{\text{nd}}$~position (March,~2021).}.

A preliminary stage is performed to join all the properties of each entity
variation. To do this, a simple MapReduce operation is performed
over the database assuming that properties are encoded using the
\emph{flattened object-key pattern}. Spark~\cite{spark-web} was used to
implement this stage. First, every database pair is mapped to a new pair
whose key is the name of the entity type along with its identifier, and the
value is formed by the property's name and its type. Then, the
\emph{reduce} operation joins all pairs of objects that belong to the same
object. The result is a set of JSON objects that are similar to those
stored in a document database. Now, the two stages of the common strategy
are performed: a MapReduce processing to obtain the set of variation
schemas, followed by the generation of the \uschema{} model, which is
similar to the document model with the exception of the key generation
using the rule R7.

The schema extracted for the running example is the same as for documents,
shown in Figure~\ref{fig:schema_doc_kv_columnar1}. The union schema is
shown in Figure~\ref{fig:union_schema_doc_kv_columnar}.

The schema extraction process was validated using a database built for the
running example, as well as using a real-world dataset. The same
\emph{EveryPolitician} dataset used with MongoDB
(Section~\ref{sec:document-mapping-implementation}) was inserted into
Redis.

The performance of the Redis schema inference process implementation versus
the query gets better than in MongoDB or Neo4j. This is because the query
itself has to process the whole database, as Redis does not include a query
language. Note also that in absolute times, the Redis implementation is the
slowest, which confirms that the calculation of an aggregate value is not
an appropriate operation for a K/V store. As in previous implementations,
the ratio goes down from~\numprint{11.46}x to~\numprint{5.76}x as the
database doubles.

\section{Representing Columnar Databases as \uschema{}
  Models\label{sec:columnar-inference}}


\subsection{A Data Model for Columnar Databases\label{columnar-datamodel}}

\begin{figure*}[!htbp]
  \includegraphics[width=\textwidth]{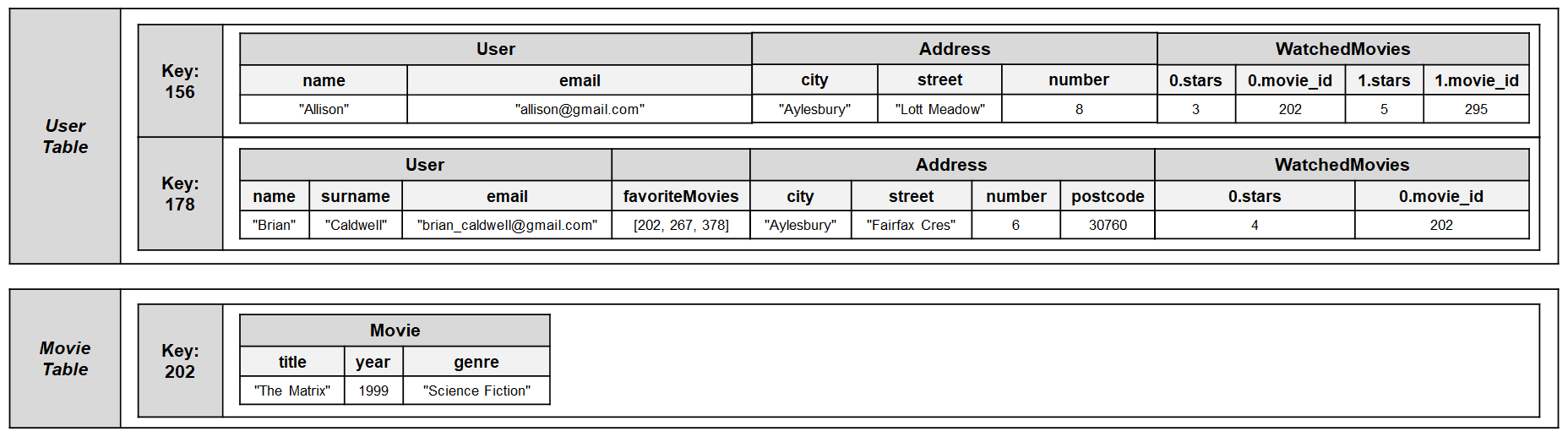}
  \caption{Columnar database example for the running
    example.\label{fig:columnar-example}}
\end{figure*}

In columnar databases, data is structured in a similar way to relational
databases. In the most popular columnar databases (Hbase~\cite{hbase-web}
and Cassandra~\cite{cassandra-web}),\footnote{This can be observed in
  \url{https://db-engines.com/en/ranking}. Cassandra appears in
  the~10$^{\text{nd}}$~position and HBase in the~22$^{\text{nd}}$~position
  as of March,~2021.} a \emph{database} or \emph{Keyspace} schema $S$ is
composed of a set of tables $T=\{t_i\}, i=1..n$, and each table $t_i$
usually stores data of a single entity type. As in relational databases,
each table has a name, and is organized in rows and columns, but rows have
a more complex structure than in relational tables because they are
organized in column families. A table $t$ is therefore defined in terms of
a set of \emph{column families} $F^t=\{F_j^t\}, j=1..m$.
Moreover, each row $r$ belonging to a table $t$ contains a \emph{row key}.
Figure~\ref{fig:columnar-example} shows an example of columnar database for
the running example, which has the \textit{User} and \textit{Movie} tables.
The \textit{User} table contains three column families: \textit{User},
\textit{Address}, and \textit{WatchedMovies}. The \textit{Address} and
\textit{WatchedMovies} relationships of the running example are represented
as column families, and the \emph{FavoriteMovies} relationship is
represented as a column of the \textit{User} family, which records an array
of references to \textit{Movie}. In the case of Cassandra, column families
will be equivalent to {\em User Defined Types} (UDTs): in a Cassandra
table, the type of an attribute can be either a predefined type or a UDT.
Thus, the \textit{User} table could have the four attributes: \textit{name}
and \textit{email} whose type would be \textit{Text}, and \textit{address}
and \textit{watchedMovies} whose types would be the UDTs \textit{Address}
and \textit{Movie}, respectively.

Columnar databases also record semi-structured data, and they are normally
schemaless, which means that structural variation is possible: the set of
columns present for each column family can vary in different rows. In the
Figure~\ref{fig:columnar-example}, the structure of the \textit{Address}
object is different for each of the two \textit{User} objects; moreover,
the second row has an additional \textit{surname} column for the
\textit{User} column family.

We will suppose that a table has a \emph{default} column family that
includes the attributes of the root entity type that corresponds to the
table. The rest of column families represent aggregated entity types.
(Again, in the case of Cassandra, the set of attributes in the table that
are not UDTs will form the default column family.) In the example, the
default column family is \textit{User}, with \textit{Address} and
\textit{WatchedMovies} as aggregated entities. Note that
\textit{WatchedMovies} aggregates an array of objects, so the name of the
columns is formed by using the \emph{flattened object-key
  pattern}\footnote{\url{https://hbase.apache.org/book.html#schema.casestudies.custorder.obj.denorm}.}
(``\texttt{<property>.<index>.<aggregated-property>}''), where the property
name is the name of the column family and can be omitted. For example:
``{\tt 0.stars}'', ``{\tt 0.movie\_id}'' in
Figure~\ref{fig:columnar-example}.

As column families are considered a way of embedding objects into a root
object, the data model defined for Key-Value and Document stores is
applicable for columnar stores, that is, the \emph{aggregate-oriented data
  model}.

\subsection{Canonical mapping between Columnar Databases and \uschema{}
  Models\label{sec:columnar-mapping}}

In the case of columnar stores, the canonical mapping would be the same as
the one defined for document stores. Relationship type would be
the only element of \uschema{} not included in the columnar model.

\subsection{Reverse Mapping Completeness}

The data model for columnar databases includes the same abstractions than
those established for the document data model. Thus, the reverse mapping
rules are the same to those introduced in Section~\ref{reverse-document}.
Only relationship types do not have a direct mapping to the model, and the
same approach used in documents can be implemented: the new entity type
with a name convention to hold the structure residing in the references,
and the reference itself on the origin entity type variations.

\subsection{Implementation and Validation of the Forward Mapping for
  HBase and Cassandra\label{sec:hbase-implementation}}

We implemented the forward mapping for Hbase and Cassandra. In the case of
HBase, we applied the common strategy of
Section~\ref{common-implementation}, with the MapReduce operation
identifying the default and aggregated column families, and building the
variation schemas. In the case of Cassandra, the API was used to retrieve
the database schema, and then build the \uschema{} model.


Validation was carried out as described in Section~\ref{common-validation}.
A database was created for the running example, and the same
\emph{EveryPolitician} real world dataset used in MongoDB and Redis,
introduced in Section~\ref{sec:document-mapping-implementation}, was
injected into Hbase and Cassandra.

Figures~\ref{fig:schema_doc_kv_columnar1}
and~\ref{fig:union_schema_doc_kv_columnar} show the variation schemas
obtained for the running example, which are the same for all the
aggregation-based stores.

As shown in Table~\ref{tab:database-times} and
Figure~\ref{fig:inference-ratio}, Hbase shows the best performance of the
inference regarding the ratio relative to the aggregated query. As with all
systems, with a slight difference in the two bigger databases
(\numprint{1.8}x to~\numprint{2.04}x), the ratio decreases as the size of
the database increases. This confirm the scalability of the schema
extraction approach. HBase, like Redis, is specialized in fast
random-access queries, but the aggregated query has to process most of the
database, making the times very close to the full process of the database
performed in the schema extraction. Thus, ratios go from just~6x slower to
around~2x slower in the case of the Larger databases.

The performance of building the Cassandra model was not recorded as no
inference process is required because the schema is already declared.

\section{Representing Relational Databases as \uschema{}
  Models\label{sec:relational-model}}


\subsection{The Relational Data Model\label{relational-datamodel}}



Unlike NoSQL logical data models, there exists a standard relational data
model which is formally defined through relational algebra and calculus.
Being ``schema-on-write'' is another significant feature that
differentiates relational databases from NoSQL stores: schemas must be
declared prior to store data in tables. The relational model is based on
the mathematical concept of \emph{relation} and its representation in form
of \emph{tables}~\cite{codd-1970}. A detailed description of the relational
model can be found in~\cite{datamodels-1982,elmasri-2015}.

\begin{figure*}[!htb]
  \centering
  \includegraphics[width=0.9\textwidth]{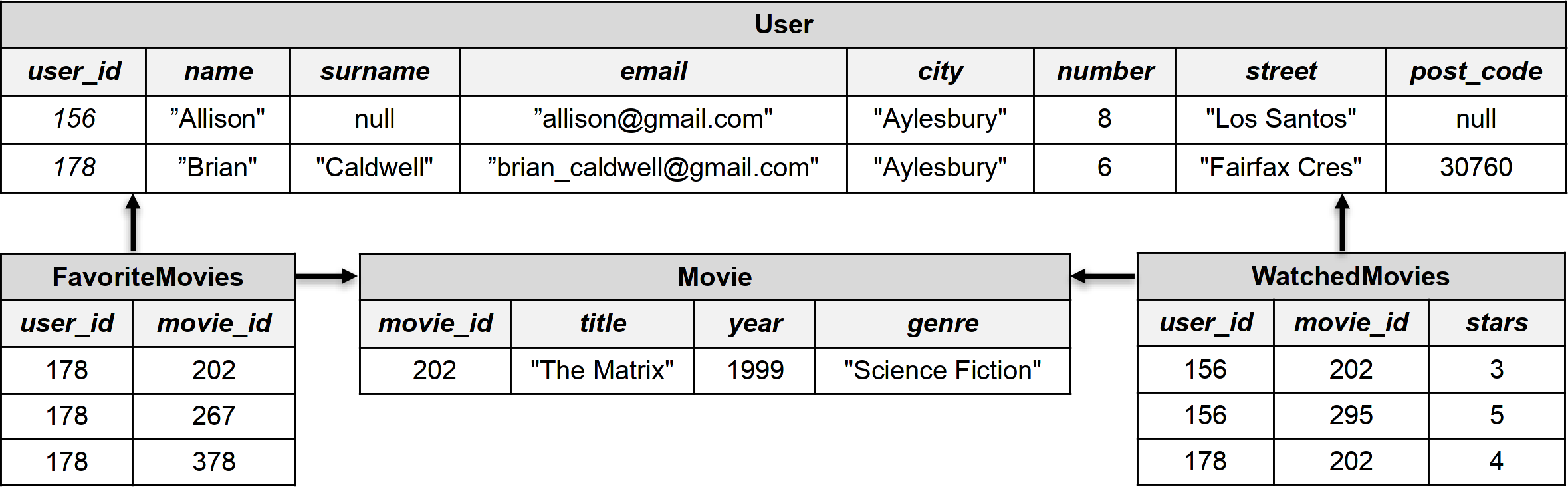}
  \caption{``User Profile'' relational example.\label{fig:relation-example}}
\end{figure*}

A relational schema consists of a set of relation schemas. Each relation
schema specifies the relation name, the attribute names and the domain
(i.e.,~type) of each attribute. Relationships between relations are
implicitly represented by key propagation from a relation schema to another
(one-to-one and one-to-may relationships) or either by a separated relation
schema (many-to-many relationships). Therefore, relation schemas can
represent entity types or relationship types. A relational schema is
instantiated by adding tuples to each relation. Each relation has one or
more attributes that form the key (\textit{primary key}), and each tuple is
uniquely identified by the values of the key attributes. Relations are
represented as \emph{tables}, and the term \emph{column} is used to refer
to the attributes, while \emph{rows} name the tuples of a relation. A table
can declare \emph{foreign keys}: one or more columns that reference to the
primary key of another table in a key propagation.
Figure~\ref{fig:relation-example} shows a relational database example for
the schema of the running example. \textit{User} and \textit{Movie} tables
represent the entity types of identical name, \textit{WatchedMovies} and
\textit{FavoriteMovies} tables represent the many-to-many relationships
from \textit{User} and \textit{Movie} in the conceptual schema of the
running example, and \textit{User} aggregates \textit{Address} by
incorporating its attributes. Note that \textit{Address} could be a
separate table related by foreign key, but it has been integrated into
\textit{User} because they hold a one-to-one relationship.

In the last four decades, conceptual and logical schemas for relational
systems have been extensively studied, and a lot of methods and tools are
available for using them in the whole database life cycle.
Entity-Relationship (ER)~\cite{datamodels-1982}, Extended ER
(EER)~\cite{elmasri-2015} and Object-Orientation modeling are the most
widely used formalisms to model conceptual and logical schemas for
relational databases. As explained in Section~\ref{sec:uschema-concepts},
the main concepts of such formalisms are included in \uschema{}, which
redefines most of them, and adds some other concepts.




\subsection{Canonical Mapping between Relational Model and \uschema{}
  Models\label{sec:relational-mapping}}

The relational model is completely integrated in \uschema{}, but the latter
has the \texttt{Aggregate} element which is not present in relational
schemas. Moreover, all the tuples have the same structure, so that the
number of structural variations for an entity type is limited to one. Next,
we expose a set of rules that specify the canonical mapping between
relational and \uschema{} models. We will use the terminology of table data
models.



%

\urule{R1}. A relational schema $D$ corresponds to a \texttt{uSchemaModel}
instance $uS$ in \uschema{} (i.e., a \uschema{} model) with the same name:
\[uS \leftrightarrow R\, \|\, \{uS.name = name(R) \}\]


\urule{R2}. Each table $t$ in $R$ representing an entity type maps to two
elements of $uS$: a root \texttt{EntityType} $et$ with the same name, and a
\texttt{StructuralVariation} $sv$ that represents the only structure of the
table that exists in the database. An identifier is generated for the
variation $sv$ and its features are mapped to the columns of the table $t$
by applying the rules R4 to~R6. This mapping can be expressed as follows:
\[\begin{aligned}%
et \leftrightarrow t &\, \|\, \{et.name = name(t) , \, et.root \leftarrow true\},\\
sv \leftrightarrow t &\, \|\, \{sv.id \leftarrow idgen(),\, sv.features \leftrightarrow
    columns(t) \}
\end{aligned}
\]
\texttt{EntityType} instances are included in $uS.entities$
and $sv$ is included in $et.variations$.


\urule{R3}. Each table $r$ in $R$ representing a relationship type maps to
two elements of $uS$: a \texttt{RelationshipType} $rt$ with the same name,
and a \texttt{StructuralVariation} $sv$ that represents the only structure
of the table that exists in the database. The mapping between $r$ and $sv$
is solved as in rule R2.
\[\begin{aligned}
    rt \leftrightarrow r & \,\|\, \{rt.name = name(r)\}, \\
    sv \leftrightarrow r & \,\|\, \{sv.id \leftarrow idgen(),\, sv.features
    \leftrightarrow columns(r) \}
  \end{aligned}
\]
\texttt{RelationshipType} instances are included in $uS.relationships$ and
$sv$ is included in $et.variations$.

\urule{R4}. Each column $c$ of a table $t$ is mapped to an
\texttt{Attribute} $at$ with the same name, and the data type of the column
will map to one of types defined in the \texttt{Type} hierarchy of
\uschema{} (a mapping between types has to be specified for each relational
system.) The mapping can be expressed as follows:
\[at \leftrightarrow c \,\|\, \{at.name = name(c), \, at.type \leftrightarrow
    type(c)\}\]
Attributes of an \texttt{EntityType} $et$ are included in the collection
$sv.features$, where $sv$ is the only structural variation that $et$ has.

\urule{R5}. The primary key $pk$ of a table $t$ is mapped to a \texttt{Key}
$k$ and the collection $k.attributes$ includes the attributes that maps to
the columns that form $pk$. The name of $k$ is the name of the attribute
(if there is just one), or $t.name + "\_pk"$ otherwise (\emph{pkname()}
function):
\[\begin{aligned}
    pk \leftrightarrow k \, \| \,& \{k.name = pkname(pk),\\
&    k.attributes \leftrightarrow columns(pk)\}
\end{aligned}
\]

For each attribute $at \in k.attributes$, $at.key = k$. $k$ is also
included in $sv.features$.

\urule{R6}. Each foreign key $fk$ of a table $t$ to a table
$s = target(fk)$ is mapped to a \texttt{Reference} $rf$, and the collection
$rf.attributes$ includes the attributes that map to the columns that form
$fk$. The name of $fk$ is the name of the attribute (if there is just one),
or $s.name + "\_fk"$ otherwise (\emph{fkname()} function). The reference
$rf$ is included in $sv$. It also refers to the entity type that maps to
the target table $s$:
\[\begin{aligned}
    fk \leftrightarrow rf\,\|\, & \{rf.name = fkname(fk), \\
    &    rf.attributes \leftrightarrow columns(fk),  \\
    & rf.refsTo = map(s) \}
  \end{aligned}\]

\subsection{Reverse Mapping Completeness}

The \uschema{} elements that are not present in the relational model are
\texttt{Aggregate}, and (multiple) \texttt{StructuralVariation}. Next, we
describe some possible mappings for these elements.

\begin{itemize}

\item The canonical mapping only takes into account a
  \texttt{StructuralVariation} per schema type (resp. table). If an schema
  type has several \texttt{StructuralVariation}s, then two possible
  alternatives are:~(i)~mapping each variation to a table with a
  distinctive naming scheme, and~(ii)~mapping all variations to a single
  table where the columns result of the union of the set of properties of
  each structural variation. In the latter case, the tuples of the table
  will have \texttt{NULL} values in the columns not corresponding to their
  structural variation. Obtaining the different entity variations from a
  table would require the analysis of all the tuples to register all the
  different set of non-NULL columns. This could be carried out with a
  similar operation to the MapReduce described in the common strategy of
  Section~\ref{sec:commonstrategy}.

\item Each \texttt{Aggregate} $ag$ in an \texttt{StructuralVariation} $sv$
  of a given \texttt{EntityType} $et$ could be mapped to elements of the
  relational model also in several ways:~(i)~an additional table $t$ with
  the name of the aggregate $ag.name$ and the columns mapped to properties
  in the \texttt{StructuralVariation} $ag.aggregates$ using rules R4 to~R6.
  A foreign key column is added to $t$, and a primary key to the table
  mapped to $et$.~(ii)~If the aggregate cardinality is one-to-one, the
  attributes of the $ag.aggregates$ variation could be incorporated into
  the table that maps to $et$. The aggregation relationship between
  \textit{User} and \textit{Address} in the running example schema has been
  mapped using the second alternative, as shown in
  Figure~\ref{fig:relation-example}.


\end{itemize}

%

\subsection{Implementing and Validating the Relational Schema Extraction
  Process\label{sec:relationImpl}}

In the case of relational databases, it is not necessary to infer schemas:
\uschema{} models can be obtained from relational schema declarations. We
chose MySQL to implement the set of rules exposed above for the relational
to \uschema{} mapping.
Rule~R3 cannot be applied as the schema does not distinguish between
relationship and entity tables. This information could be provided, for
example, through name conventions, which could also be used to specify
aggregation tables.

The model generation process is straightforward, and it works following the
described mapping rules. First, R1 is applied to create and name the model,
then an \texttt{EntityType} and a \texttt{StructuralVariation} are created
for each table~(R2). An \texttt{Attribute} is created for each column of a
table~(R4). Next, \texttt{Key}s are created for primary keys in tables,
which will have references to \texttt{Attribute}s that have been
instantiated previously for the columns that are part of the primary
key~(R5). Finally, \texttt{Reference}s are created for foreign keys in
tables, and each \texttt{Reference} will be connected to elements
previously created according to the \uschema{} metamodel~(R6).


The validation has been performed on the \emph{Sakila} database available
at the MySQL official website.\footnote{Sakila can be downloaded from
  \url{https://dev.mysql.com/doc/index-other.html}, and documentation is
  available at
  \url{https://dev.mysql.com/doc/sakila/en/sakila-structure.html}.} Sakila
contains~16~tables, and the average numbers of columns and references
between tables are, respectively,~\numprint{5.6} and~\numprint{1.4}. The
smallest table has~3 columns, and the biggest one~13~columns. We have
checked the correction of the \uschema{} model generation by comparing the
model obtained with the information on the database available at the MySQL
website (SQL creation files and official diagrams). In the study of
performance and scalability, as with Cassandra, relational databases have
not been considered because schemas are already available.

\section{Related work\label{sec:relatedwork}}

In this section, the \uschema{} metamodel will be contrasted to some
relevant generic metamodels defined for database schemas, and the schema
inference strategy to others published for NoSQL stores.

\subsection{Generic Metamodels}

\textbf{DB-Main} was a long-term project aimed at tackling the problems
related to database evolution~\cite{hick2003,hainaut1994}. The DB-Main
approach was based on three main elements:~(i)~The Generic
Entity/Relationship (GER) metamodel to achieve
platform-independence;~(ii)~A transformational approach to implement
operations such as reverse and forward engineering, and schema mappings;
and~(iii)~A history list to record the schema changes~\cite{hick2003}.
Here, our interest is focused on the two former elements. The generic GER
metamodel was defined as an extension of the ER
metamodel~\cite{datamodels-1982}. Conceptual, logical, and physical models
could be represented in GER. Models for a particular paradigm, system, or
methodology were obtained by means of~(i)~selecting necessary GER
elements,~(ii)~defining structural predicates to establish legal assemblies
of that elements, and~(iii)~choosing an appropriate visual diagram.
Regarding schema transformations, a set of basic transformations were
defined, and the signature of each of them (name, input, and output) was
specified in a particular format to be used to record changes in the
history list. Our proposal differs of the GER approach in several
significant aspects.

\paragraph{\textit{A.} Support of semi-structured data in NoSQL stores}
It is convenient to remark that our approach shares objectives with
DB-Main. However, DB-Main was focused mainly on relational systems, and
also on earlier database systems. Instead, we are interested in both
structured and semi-structured data, specially in the emerging NoSQL stores
and relational databases.

\paragraph{\textit{B.} Physical and conceptual level separation in
  different metamodels} \uschema{} is intended to represent logical
schemas, so that conceptual and physical schemas are separately modeled.
Instead of mixing all the information in a single metamodel, we have
considered more convenient to separate the large amount of physical
concepts in their own metamodel and to have a simpler conceptual model.
Because of this concern separation, reusability is promoted, and models are
kept simple and readable. The conceptual and physical metamodels are out of
the scope of this paper. At this moment, we have defined a physical
metamodel for MongoDB, as described in~\cite{pablo-comonos2020}.

\paragraph{\textit{C.} Concrete schemas are directly represented in
  \uschema{}} Unlike GER, we do not have to define a sub-model of
\uschema{} for each database system. \uschema{} acts as a pivot
representation, able to represent NoSQL and relational schemas for all
paradigms. The set of rules that maps each data model to \uschema{}
determines the \uschema{} elements involved, and therefore the valid
structures.

\paragraph{\textit{D.} Structural variation representation}
A central notion of \uschema{} is structural variation. Variations of
entity and relationship types can be represented. This information can be
useful in different tasks. For example, variations are used to identify
whether an entity type contains a type hierarchy (e.g.~a \emph{Product}
hierarchy) in~\cite{alberto-subtypes2020}. Variations also allow to analyze
the database evolution, or can be used to generate test datasets, among
other tasks.

\paragraph{\textit{E.} Solution based on MDE specification}
As indicated in Section~\ref{sec:uschema-model}, we have defined \uschema{}
with the Ecore metamodeling language with the purpose of taking advantage
of MDE technology integrated in the EMF framework~\cite{steinberg-emf2009}.

\paragraph{\textit{F.} Schema extraction}
In DB-Main, a different schema extractor had to be developed for each
database system. In our case, a common strategy have been defined which
address the scalability and performance issues.


\textbf{Model Management} (\textbf{MM}) is an approach aimed to solve
\textit{data programmability} problems which normally involve complex
mappings between data schemas of different
sources~\cite{bernstein2007b,bernstein2000}. A set of operators between
models are proposed, such as \emph{match}, \emph{union}, \emph{merge},
\emph{diff}, or the \emph{modelgen} operator that generates a schema from
another. In~\cite{bernstein2007b}, building a universal metamodel is
considered a feasible way of developing tools to specify mappings, although
it does not seems the more adequate alternative because of the large
complexity of the required metamodel.

Two universal metamodels for applying Model Management are presented
in~\cite{atzeni2009} and~\cite{kensche2007}. \textbf{Paolo Atzeni et
  al.}~\cite{atzeni2009} described a universal metamodel based on a
three-level architecture similar to those defined in the EMF framework and
used in our work: a metamodeling language (Ecore) is used to define
metamodels (\uschema{} in our case), which, in turn, are used to create
models (schemas in our case).
In~\cite{atzeni2009}, a set of~13~meta-constructs were defined to represent
the concepts used in different data formalisms. For example,
\emph{Abstract} is proposed to model autonomous concepts such as ER
entities or OO classes, \emph{AbstractAttribute} to model references, and
\emph{Generalization} to model \emph{Abstract} hierarchies. This proposal
overlooked the already existing MDE frameworks, in particular EMF/Ecore.
Instead, the authors started from scratch, and they even proposed a
dictionary structure to store models as instances of the universal
metamodel. Schemas are expressed by indicating, for each element, the
construct at the level of the data model from which is instantiated, and
for each of these constructs its meta-construct at the level of the
universal metamodel.
The metamodel was accompanied by a basic tooling for textual and graphical
visualization.

The main differences between this universal metamodel and \uschema{} are
the following:

\paragraph{\textit{A.} A different purpose and meta-modeling architecture}
While the universal metamodel of Atzeni et al. is aimed to instantiate data
models, \uschema{} is a unified metamodel able to represent schemas of a
variety of databases. Therefore, the metamodeling architectures are
different: Universal metamodel/Data Model/Database Schemas vs.
Ecore/\uschema{}/Database Schemas.
It is worth noting that our approach does not prevent the definition of
metamodels for representing any existing data model that is integrated in
\uschema{}. However, as indicated in Section~\ref{common-implementation},
we have considered that creating these metamodels would not provide any
benefit as intermediate representation, as the variation schema to data
model transformations would be very close to the variation schema to
\uschema{} transformation.


\paragraph{\textit{B.} Availability of tools for basic model operations}
While we used the EMF metamodeling architecture to create \uschema{},
Atzeni et al. had to implement their own metamodeling architecture from
scratch, as well reporting and visualization tools. Instead, EMF provide
tools supporting model comparison (EMF
Compare)\footnote{\url{https://www.eclipse.org/emf/compare/}.} and model
diff/merge operations (EMF
Diff/Merge),\footnote{\url{https://wiki.eclipse.org/EMF_DiffMerge}.} as
well as model-transformation languages to implement the \emph{modelgen}
operator.

\paragraph{\textit{C.} Relationship types and structural variations}
The expressiveness of the Universal metamodel is covered by \uschema{}
elements. In addition, \uschema{} includes the notions of relationship
types and structural variations, which are convenient to represent schemas
of NoSQL stores.

\textbf{GeRoMe} is another generic metamodel proposed for Model
Management~\cite{kensche2007}. A \emph{role-based modeling\/} is applied to
define a metamodel able to represent different data models. In
mid-nineties, role-based modeling approaches received attention in the
context of object-oriented programming to model the multiple-classification
and object collaborations~\cite{reenskaug1996}. However, that interest has
decreased over the years because languages and tools do not support the
notion of \emph{role}. Extended ER, Relational, OWL-DL, XML Schema, and UML
were analyzed in GeRoMe with the aim to identify their similarities and
differences. Then, a set of roles was established, and the role-based
metamodel created. \uschema{} clearly differs of GeRoMe in its purpose and
the kind of representation of the generic metamodel. Our unified metamodel
has been defined by applying object-oriented conceptual modeling, the
technique commonly used currently to create metamodels, and using a
well-know metamodeling architecture.

As far as we know, neither of the three generic metamodels here considered
(GER, Atzeni et al., and GeRoMe) has evolved to include elements specific
of NoSQL stores. Therefore, none of them has addressed the representation
of structural variations or relationship types. In the case of DB-Main, the
tool can currently be acquired from the Rever
company\footnote{\url{https://www.dataengineers.eu/en/db-main/}.} as a tool
to simplify data engineering tasks.

More recently, several metamodels have been proposed to represent NoSQL
schemas. \textbf{SOS} is a metamodel designed to represent schemas of
aggregate-based stores~\cite{atzeni2012}. With this uniform representation,
a NoSQL schema consists of a set of collections (\emph{Set} metaclass),
which can contain \emph{Struct}s and \emph{Attribute}s. An \emph{Attribute}
represents a key-value property, and a group of key-value pairs is modeled
as a \emph{Struct}. \emph{Struct} and \emph{Set} can be nested. Later, SOS
evolved to the \textbf{NoAM} (NoSQL Abstract Model) metamodel, which was
defined as part of a design method for aggregate-based NoSQL
databases~\cite{atzeni2016,atzeni-2020}. NoAM was designed as an intermediate
representation to transform aggregate objects of database applications into
NoSQL data. A NoAM database is a set of collections that contains a set of
blocks. A block contains a set of key-value pairs, and each block is
uniquely identified. In~\cite{atzeni2016}, several strategies are described
to represent a collection of aggregate objects in form of a NoAM database.

NoAM and SOS were designed with a purpose different to \uschema{}. SOS aims
to achieve a uniform accessing, and NoAM is part of a design method.
Instead, \uschema{} has been devised to have a uniform representation able
to capture data models of NoSQL and relational data models, with the aim of
facilitating the building of database tools supporting several database
systems. Therefore, \uschema{} offers a higher level of abstraction than
SOS and NoAM. These representations are closer to the physical level than
the logical. Thus, some key aspects for a logical schema are neglected,
such as the relationships between entities. In addition, the existence of
structural variations is not considered. Finally, MDE technology was not
used in their definition.

\emph{ERwin Unified Data Modeler (ModelSet)\/} is a project outlined in an
article in infoQ~\cite{allenwang2016} whose purpose is very close to ours.
However, to our knowledge, results of that project have not been published
yet. In~\cite{allenwang2016}, Allen Wang, responsible of the ERwin project,
pointed out on the importance of ``using logical models describing business
requirements and de-normalizing schema to physical data models'' in
database design. A simple unified logical model is shown to represent three
kind of schemas: columnar, document, and relational schema. The metamodel
only includes four elements. The three basic modeling constructs:
\emph{Entity}, \emph{Relationships}, \emph{Properties} (of entities),
and \emph{Tags} are used to add additional information to basic constructs.
A physical model should be built for each system. Query and data production
patterns are defined on the logical model for its transformation into
physical model. Several significant differences are found between
\uschema{} and the ERwin metamodel:~(i)~\uschema{} is not only able to
represent aggregate-based systems, but also graph stores;~(ii)~\uschema{}
is more expressive, ModelSet only includes the three basic constructs of
modeling, but this is similar to our variation schemas that are input to
the analysis process;~(iii)~Being \uschema{} a representation at higher
level of abstraction, the definition and implementation of operations such
as schema mapping, visualization, or schema discovery are
easier;~(iv)~\uschema{} represents structural variations;~(v)~Instead of a
proprietary tool, \uschema{} is part of a free data modeling tool.

The Typhon project\footnote{\url{https://www.typhon-project.org/}.} is an
European project aimed to create a methodology and tooling to design and
develop solutions for polystore database systems. As part of this project,
the TyphonML~\cite{TyphonML-TR-2018} language has been built, which allows
schemas to be defined in a database system-independent way. Columnar,
document, key-value, graph, and relational schemas can be defined with
TyphonML. Typhon schemas can also express mappings from schemas to the
physical representation.
Some remarkable differences with \uschema{} are:~(i)~TyphonML is not a
language defined on a unified metamodel, instead \uschema{} is separated
from any schema declaration language. In fact, we have created the Athena
language on \uschema{}, and other languages could be
defined\footnote{\url{https://catedrasaes-umu.github.io/NoSQLDataEngineering/tools.html}.};~(ii)~The
existence of structural variation in NoSQL systems is not
considered;~(iii)~As can be observed in~\cite{TyphonML-TR-2018}, for each
paradigm, the TyphonML metamodel includes logical and physical aspects;
Instead, our choice is to separate both levels of abstraction in two
metamodels as pointed out in~\cite{pablo-comonos2020};~(iv)~Although graph
stores are represented, the concept of relationship type is not included in
TyphonML;~(v)~Aggregates are not represented as a separate concept, but the
same metaclass \emph{Reference} represents both aggregates and references
by using the boolean attribute \emph{isComposite} to record the kind of
relationship; Instead \uschema{} represents aggregates and references with
two metaclasses, which allows us to have a complete semantics. The logical
elements of TyphonML are limited to \emph{Entities} that aggregate
\emph{Attributes} and \emph{References}, while our unified metamodel has a
wider and richer set of semantic concepts.

A~2-step model transformation chain aimed to transform conceptual models
into physical schemas (Cassandra stores are considered) is described
in~\cite{abdelhedi2017}. Logical models are generated in an intermediate
step. Conceptual schemas and logical schemas are represented by means of
very simple metamodels. A conceptual schema is formed by a set of classes
and datatypes, and classes include attributes and relationships. A
relational model is used as logical model, to which relationships are
added. This proposal has some flaws such as~(i)~relationship types and
references are not distinguished, which is necessary for graph
schemas,~(ii)~references have not properties, and~(iii)~the separation
between logical and conceptual model is not justified because they include
the same concepts but different names; this can be observed in the very
simple conceptual-to-logical transformation shown in the paper.


Table~\ref{table:metamodel-approaches} summarizes the discussion made
above, and compares the generics metamodels considered according to several
criteria.

\begin{sidewaystable*}[htb!]
\begin{tiny}
\noindent%
\begin{tabular}{p{.1\textwidth}p{.1\textwidth}p{.12\textwidth}p{.12\textwidth}p{.1\textwidth}p{.12\textwidth}p{.14\textwidth}}
\toprule
     &  \textbf{DB-Main} & \textbf{Universal Metamodel}                         &  \textbf{GeRoMe}                       & \textbf{SOS/NOAM}             &  \textbf{ERwin Unified Data Modeler (ModelSet)} &  \textbf{U-Schema}                   \\

\midrule

\textbf{Aim}            & Evolution tool          & Model Management                                                                                    & Model Management                    & \raggedright Uniform access / Database Design               & Modeling tool      & Database engineering toolkit                                                                                                                                                                              \\


\textbf{Supported Databases Paradigms} & Relational, OO, ER and Early databases                        & Any metadata formalism and database schema         & Any metadata formalism and database schema                                                 & NoSQL                                                                    databases & Relational, document, and columnar                                                        & Relational and NoSQL (columnar, document, graph, and key-value)                                                                                                                                                 \\

\tabularnewline


\textbf{Unified Metamodel}                                                                         & GER based on ER                                             & Set of 13 meta-constructs to cover all data models
& Set of 48 roles to cover all data models
& Collection, Struct or Block, and Attribute (very near to physical model) & Entity, Property, and Relationship & Set of concepts to cover NoSQL and relational schemas
\\

\tabularnewline


\textbf{Metamodeling Language}       & From scratch         & From scratch                                                                              & From scratch        &    N/A              & From scratch              & Ecore (Eclipse Modeling Framework, EMF)   \\

\tabularnewline


\textbf{Levels of metamodeling}      & GER metamodel and restrictions to define data model         & \raggedright SuperMetamodel/Data Model/Schema (Own architecture)                 & Use of role-based modeling                                         &     Abstract metamodel / Instances  & Unified metamodel / Models (schemas of data models)               & Ecore / U-Schema / Models (schemas of a data model)                        \\

\tabularnewline


\textbf{Defining concrete schemas}   & Selection of GER elements and definition of constraints     & Model elements are annotated with meta-construct to which belong it                           & Model elements play one of their roles                 &       Programmatically, Java instances                   & Instances of the metamodel (proprietary solution)                                         & Instances of the metamodel (use of mapping rules)                                                                                                                                                         \\

\tabularnewline


\textbf{Schema Levels}               & Conceptual, Logical and Physical                & Conceptual and Logical                                                   & Conceptual and Logical                           &  Very simple uniform representation                     & Conceptual and Logical (Physical separated)            & Logical (Conceptual and Physical separated)       \\

\tabularnewline


\raggedright \textbf{Schemaless supported}  & Not addressed              & Not addressed                                            & Not addressed                          &    Structural variations are supported, but not extracted or represented               & Not addressed                 & Structural variations are modeled                                     \\

\tabularnewline


\textbf{Output}         & ER diagrams and text            & ER diagrams and text                                            &      Own visualization         &      N/A              & Unified ModelSet notation                           & \uschema{} models in form of Neo4j graphs  \\

\tabularnewline


\textbf{Schema extraction}          & A schema extractor for each system                         & Not addressed            & Not addressed            &         N/A            & No details provided, except the use of machine leaning and statistics are obtained     & \raggedright Common strategy of 2 steps: MapReduce and \uschema{} model building process         \tabularnewline

\textbf{Scalability and Performance}          & Not addressed                        & Not addressed             & Not addressed            &         N/A            & No details  provided  & MapReduce operation on NoSQL stores                                                                                                                                                \\

\bottomrule
\end{tabular}
\end{tiny}
\caption{Approaches defining a Generic Metamodel.\label{table:metamodel-approaches}}
\end{sidewaystable*}

\subsection{NoSQL Schema Extraction Strategies}

Recently, several approaches to extract schemas from NoSQL document stores
have been
published~\cite{wang-schema2015,klettke-schema2015,colazzo-datasets2012}. A
detailed study of these works can be found in~\cite{severino2017} where
they were contrasted to our previous approach for document
stores~\cite{sevilla-er2015}. Moreover, some works on schema extraction
from JSON datasets have also been presented.
In~\cite{colazzo-datasets2012}, a MapReduce is used to obtain a collection
of key-value pairs from an input JSON dataset. In each pair, the key is a
document specifying the structure or type of a JSON object in the dataset,
and the value records the number of elements of the that type. In a second
step, heuristics are applied to merge similar types.

The main differences of the approach described here with previous
extraction strategies are the following:~(i)~They are focused on document
stores. Instead, we have defined a general strategy applicable to the four
main NoSQL paradigms and the relational
model;~(ii)~Like~\cite{colazzo-datasets2012}, our approach use a MapReduce
operation to improve the efficiency of the inference process;~(iii)~Our
inference strategy discovers relationships between entities, and structural
variations of entities.~(iv)~The output of our inference process is a model
that conforms to an Ecore unified metamodel. In this way, we can take
advantage of benefits offered by MDE, which were commented in
Section~\ref{sec:uschema-metamodel}.


An MDE-based reverse engineering approach for extracting conceptual graph
schemas is described in~\cite{comyn-wattiau2017}. \texttt{CREATE} Cypher
statements are analyzed to obtain a graph model: a graph is formed by nodes
and edges, and nodes have incoming and outgoing edges. Then, a
model-to-model transformation generates an Extended Entity-Relation (EER)
conceptual schema model, whose elements are entities, relationships,
attributes, and IS-A relationships. Our inference process differs of this
strategy in several significant aspects, apart from being a generic
strategy and use a MapReduce to considerably improve the efficiency:~(i)~We
access stored data instead of using \texttt{CREATE} statements, which might
not be available.~(ii)~Instead of building a graph model, we create a table
with all relationships as input to the MapReduce operation; this table
allows us to obtain the cardinality of each relationship%
~(iii)~We obtain a logical schema that include structural
variations for relationships and entities.

\subsection{Data Modeling Tools}

%


With the emergence of NoSQL systems, multi-paradigm data modeling
commercial tools have proliferated. In our study of some of the most
relevant of these tools, we have found no evidence showing the use of a
unified metamodel. Next, we contrast features of these tools with those
considered in our \uschema{} approach.

These tools can be classified in two categories. A first category are
existing tools for relational databases which are incorporating some NoSQL
systems. At this moment, these tools have only added support for document
systems, being MongoDB the system integrated in the most of them.

For example, ER/Studio~\cite{erstudio-web} and ERwin~\cite{erwin-web}
provide utilities to extract and visualize schemas for MongoDB and CouchDB
since~2015. They extract schemas as a set of entity types whose properties
are the union of all fields discovered in objects of that entity, but
variations and relationships are not addressed. Recently, ERwin Data
Modeler provides an integrated view of conceptual, logical and physical
data models to help stakeholders understand data structures and meaning.

The second group is formed by new tools developed with the purpose of
offering data modeling for polyglot persistence. As far as we know,
Hackolade~\cite{hackolade-web} is the only tool that integrates database
systems for the four most common NoSQL paradigms as well as a wide number
of relational systems and other leading data technologies. Recently, it has
been announced the creation of a unified model named “Polyglot Data Model”
but no details have been published. Unlike \uschema{}, Hackolade does not
address variation and references in the NoSQL schema extraction. Entities
extracted are represented as the union of all the fields discovered in
different variations of the entity. The collision of fields with the same
name but different type is not considered but that modeler should make a
decision.


DBSchema~\cite{dbschema-web} is a tool similar to Hackolade: It allows the
developer to define schemas with a graphical layout, but also to apply a
reverse engineering process to an existing database in order to extract the
schema, as long as there is a JDBC Java driver for it. Queries can be
created in an intuitive way or either using SQL. In this tool, variations
are not considered at all, since it applies a SQL approach to infer the
schema, in which variations are not taken into account.

\section{Applications of the \uschema{} Metamodel\label{sec:use-cases}}

The usefulness of generic metamodels is well known, and has been
extensively discussed in the database literature for more than~30~years. In
this section, we shall show how \uschema{} metamodel favors the
implementation of some database operations that involve SQL and NoSQL
systems. We will present a migration process and discuss how \uschema{}
facilitates the development of database migrations, generic solutions to
query schemas, generate synthetic data for testing purposes, and visualize
schemas.


\subsection{Database Migrations\label{sec:migrating}}

Database migration is a typical task in which a unified or generic
representation provides a great advantage. Given a set of $m$ database
systems, the total number of migrators required is $m+m$ instead of
$m \times (m - 1)$. 
Here we will describe how \uschema{} models can be used to help migrate
databases when the source and target systems are different.

%


To perform a migration, the source and destination databases have to be
specified, as well the mapping rules that determine how source data are
moved to the destination database. A migration tool usually has to read all
the data in the original database, perform some processing, and write the
resulting data in the destination database. These steps can be carried out
in different ways, that can be simplified by using \uschema{} models, as
they contain all the information of entities, attributes, and
relationships. Therefore, the \uschema{} model has to be obtained prior to
the aforementioned steps.

There are several options when reading the original data. A set of queries
could be constructed to extract the data guided by its structure (i.e., its
schema). The inferred \uschema{} model from the source database can be used
to automatically generate those queries. The queries can produce a set of
interchange format files (e.g. JSON or CSV) or can act as a source feed for
a streaming process. Likewise, \uschema{} models could automate the data
ingestion procedure using bulk insertion utilities from files, generated
insert queries, or even help to build the ingestion as the last stage of a
streaming process.

The next step is to specify and execute the mapping rules between source
and destination elements. The mapping rules introduced in
sections~\ref{sec:graph-database} to~\ref{sec:relational-model}
should be adapted to the specificities of the migration. For example, an
alternate mapping could be devised for characteristics not present in the
destination data model, as was the case we showed with aggregates in a
graph data model in Section~\ref{sec:reverse-mapping-graph}. The migration
rules could be hardcoded, or either specified with a ad-hoc language. This
language would be defined taking into account the abstractions of
\uschema{}. The migration rules would include the \uschema{} source
element, the target data model element, and the mappings between the parts
that constitute the source and target elements, similarly to how we
expressed the canonical mappings before.

\subsection{Definition of a Generic Schema Query
  Language\label{schemaquery}}

Schema query languages help developers to inspect and understand large and
complex schemas. In the case of relational systems, SQL is used to query
schemas represented in form of tables in the data dictionary. In NoSQL
stores, a similar query facility is provided by some systems that require
to declare schemas, for example Cassandra~\cite{cassandra-web} and
OrientDB~\cite{orientdb-web}. In the case of schemaless NoSQL systems, the
number of variations can be very large in some domains, for
example~\numprint{21302} variations for the \textit{Company} entity type of
DBPedia are reported by Wang et al.~\cite{wang-schema2015}.
Using \uschema{}, a generic query language could be defined which would
allow querying relationships and structural variations for any kind of
NoSQL store, unlike existing solutions. As far we know, querying variations
has been only addressed in the mentioned work of Wang et
al.~\cite{wang-schema2015}, which focused on MongoDB, and only suggested a
couple of queries to illustrate the idea. A first version of our language
can be
downloaded.\footnote{\url{https://github.com/catedrasaes-umu/NoSQLDataEngineering}.}

The \uschema{} query language allows to query the schema of any type of
database system under a unique language, and even make it possible in
scenarios where the data is stored in different database systems (polyglot
persistence). Some examples of the most common queries that a developer
might need are:~(i)~get an overview of the entities and the relationships
between them,~(ii)~search variations with a set of properties,~(iii)~check
all shared properties of all variations of a specific entity. The results
of the queries could be displayed as text or a graphic representation in
the form of tables, graphs or trees (hierarchical data).

\subsection{Generation of datasets for testing
  purposes\label{datageneration}}

Automatic database generation is a point of interest in designing,
validating, and testing of research database tools and deployments of data
intensive applications. Often, researchers in the data-engineering field
lack of real-world databases with the required characteristics, or they can
not access them.

Some works have addressed the generation of synthetic data on relational
systems, and some restriction languages have been proposed to this
purpose~\cite{bruno-flexible2005,smaragdakis-scalable2009}. With
\uschema{}, a database paradigm-independent restriction language could be
defined to tailor the generation of data. In this way, a given
specification could be used to generate data for different databases. Note
that the language constructs would be at the level of abstraction of
\uschema{}, and not aligned to elements of any concrete paradigm.

%
%
This is of special importance in the case of distributed systems, as most
NoSQL deployments are. In this context, a cost model to evaluate query
efficiency is very difficult to build, given all the variables
involved~\cite{mior_nose_17}. Generating different sets of data with
different characteristics can help fine-tuning application intended
queries. For example, just changing the relationships between the entities
of a schema (for example, changing references into aggregations or vice
versa), new data that follows this change could be generated to test the
queries, helping the developers to find opportunities for optimization.

Finally, another advantage of our approach around \uschema{} is that in the
case of existing databases, their schema can be inferred into a model, and
then used to generate data that can be for the same or different
databases, matching the schema or even introducing changes, either
for performance tuning or for testing purposes.

We developed an initial version of a \uschema{}-based data generation
language with the described characteristics~\cite{alberto-comonos2020}.

%
%


\subsection{\uschema{} Schema Visualization}
\label{visualization}

When schemas are extracted they must be expressed in a graphical, textual,
or tabular format to be shown to stakeholders. Normally, they are shown as
a diagram (e.g.,~ER or UML). In a previous work, we explored the
visualization of document schemas and proposed several kinds of diagrams
for document systems~\cite{alberto-erforum2017}. Now, it is possible to
take advantage of \uschema{} to define common diagrams for logical schemas
taking into account the existence of variations if needed. Moreover,
\uschema{} could be mapped to other formats with the purpose of visualizing
schemas in existing tools.

\section{Conclusions and Future Work\label{sec:conclusions}}

Multi-platform database engineering tools commonly define a unified
metamodel to represent database schemas of a variety of systems. In this
paper, we have presented a proposal of unified metamodel that integrates
data models for relational and NoSQL systems (key-value, document, wide
column, and graph). These systems cover most of current database
applications. Our work is motivated by the growing interest in multi-model
database tooling and systems as polyglot persistence is considered
essential to satisfy needs of modern applications. With \uschema{}, we have
defined a representation able to express inferred or declared schemas at a
similar abstraction level to EER and Object-Orientation logical models.

\uschema{} is the first logical unified metamodel defined for NoSQL and
relational systems taking into account structural variation, relationship
types, aggregations, and references. Through forward and reverse mappings,
we have formally shown how \uschema{} is able to represent each considered
data model, and how \uschema{} models can be converted to schemas of the
individual models. We would like to remark that the extraction of schemas
from databases (forward mappings) have been implemented for the most widely
used NoSQL systems (Neo4j, MongoDB, Redis, Cassandra, and HBase), as well
as for one of the most used relational systems (MySQL). For each extraction
algorithm, scalability and performance were assessed. Having used the
\emph{de facto} standard Ecore to represent the schemas turns the framework
in a reusable and adaptable tool, and Eclipse modeling tooling can be used
to build database tooling.

Future work can be divided in two lines, depending on whether they have to
do with the unified metamodeling approach, or with applications based on
\uschema{}. We approached the unified representation of schemas by
separating logical and physical aspects. The metamodel presented here
concerns to the logical view, and a new metamodel will represent the
unified physical schemas. Thus, we will have \uschema{}-Physical and
\uschema{}-Logical, where physical schemas will be extracted from data
stores, and logical schemas could be directly obtained either from stores
or from physical schemas, as described in~\cite{pablo-comonos2020}.
Physical data models for each system will include data structures at
physical abstraction level, indexes, physical data distribution, among
others. Regarding improvements of \uschema{}, we will extend the
metamodel to represent constraints to support new logical validation
characteristics in some NoSQL databases, such as the MongoDB Schema
Validation. Finally, it is planned to build several tools and languages
around \uschema{}:~(i)~The migration approach outlined in
Section~\ref{sec:migrating}, which will include a \uschema{} based schema
mapping language to express specialized mappings as described in
Section~\ref{sec:uschema-mapping};~(ii)~Complete the generic Schema Query
Language introduced in Section~\ref{schemaquery};~(iii)~Define some
diagramming tools from the results in~\cite{alberto-erforum2017};~(iv)~A
universal schema definition language, that, by using \uschema{}, allows to
define schemas homogeneously through NoSQL and relational
datastores;~(v)~Support an approach of platform-independent schema
evolution through the definition of a change taxonomy implemented by a
schema change operation language; and~(vi)~This language could be used to
explore the impact of schema changes on the set of queries of an
application, following automated reinforcement learning techniques.



\bibliographystyle{plainurl}
\bibliography{bib/allstars}

\end{document}